\documentclass[10pt, prd, reprint, onecolumn, nofootinbib, showpacs]{revtex4-1}

\usepackage[utf8]{inputenc}

\usepackage{amsmath}

\usepackage{amsfonts}
\usepackage{mathrsfs}

\usepackage{amssymb}
\usepackage{units}

\usepackage{graphicx}

\usepackage{placeins}

\usepackage{verbatim}

\usepackage{accents}

\newcommand{\mrm}{\mathrm}

\makeatletter

\let\old@dmathbeg\[
\let\old@dmathend\]

\newcommand{\rovnec}[1]{\old@dmathbeg#1\old@dmathend}
\newcommand{\rovcis}[2]{\begin{equation}#1\label{#2}\end{equation}}
\newcommand{\drovcis}[2]{\begin{equation}\begin{split}#1\end{split}\label{#2}\end{equation}}
\newcommand{\drovnec}[1]{\begin{equation*}\begin{split}#1\end{split}\end{equation*}} 
\newcommand{\provcis}[1]{\begin{align}#1\end{align}}
\newcommand{\provnec}[1]{\begin{align*}#1\end{align*}}

\newcommand{\rov}{\@ifstar\rovnec\rovcis}
\newcommand{\drov}{\@ifstar\drovnec\drovcis}
\newcommand{\prov}{\@ifstar\provnec\provcis}

\newcommand{\vast}{\bBigg@{4}}
\newcommand{\Vast}{\bBigg@{5}}

\makeatother

\newcommand{\vld}{\underaccent{\tilde}}

\DeclareMathOperator{\sgn}{sgn}
\DeclareMathOperator{\diff}{d}
\newcommand{\td}{\diff\!}
\DeclareMathOperator{\diffbold}{\mathbf{d}}
\newcommand{\bd}{\diffbold\!}

\newcommand{\mbb}{\mathbb}
\newcommand{\msc}{\mathscr}
\newcommand{\mbs}{\boldsymbol}

\newcommand{\iDelta}{{\mit\Delta}}
\newcommand{\iLambda}{{\mit\Lambda}}
\newcommand{\iSigma}{{\mit\Sigma}}
\newcommand{\iXi}{{\mit\Xi}}
\newcommand{\iPhi}{{\mit\Phi}}

\DeclareFontEncoding{LGR}{}{}

\DeclareRobustCommand{\pltext}{%
  \fontencoding{T1}\selectfont\def\encodingdefault{T1}}
\DeclareRobustCommand{\textpl}[1]{\leavevmode{\pltext #1}}
\AtBeginDocument{\DeclareFontEncoding{T1}{}{}}

\DeclareMathAlphabet{\mgr}{LGR}{cmr}{m}{n}
\newcommand{\rpi}{\mgr{p}}

\newcommand{\imply}{\Longrightarrow}

\renewcommand{\[}{\left[}
\renewcommand{\]}{\right]}

\newcommand{\f}{\!\left}

\newcommand{\pder}[3][]{\frac{\partial^{#1}#2}{\partial{#3}^{#1}}}
\newcommand{\smder}[3]{\frac{\partial^2 #1}{\partial #2 \partial #3}}
\newcommand{\bpd}[2][]{\frac{\boldsymbol\partial #1}{\boldsymbol\partial #2}} 
\newcommand{\nicebpd}[2][]{\nicefrac{\boldsymbol\partial #1}{\boldsymbol\partial #2}}

\newcommand{\zrov}{{}\\{}}

\newcommand{\zrovn}{{}\\\nonumber{}}
\newcommand{\res}[2]{\left.#1\right|_{#2}}
\newcommand{\lbl}{\label}
\newcommand{\rvt}{\ .}
\newcommand{\rvc}{\ ,}
\newcommand{\rvs}{\ ;}
\newcommand{\qt}[1]{``#1''}

\newcommand{\kom}[2]{\left[#1,#2\right]}

\renewcommand{\(}{\left(}
\renewcommand{\)}{\right)}

\begin{document}

\title{Near-horizon description of extremal magnetised stationary\\ black holes and Meissner effect}

\author{Jiří Bičák}
\email{bicak@mbox.troja.mff.cuni.cz}
\affiliation{Institute of Theoretical Physics, Faculty of Mathematics and Physics,
Charles University in Prague,
V Holešovičkách 2, 180\,00 Prague 8, Czech Republic}
\affiliation{Max Planck Institute for Gravitational Physics, Albert Einstein Institute, Am Mühlenberg 1, D-14476 Golm, Germany}

\author{Filip Hejda}
\email{hejdaf@plk.mff.cuni.cz}
\affiliation{Institute of Theoretical Physics, Faculty of Mathematics and Physics,
Charles University in Prague,
V Holešovičkách 2, 180\,00 Prague 8, Czech Republic}
\affiliation{Centro Multidisciplinar de Astrofísica -- CENTRA, Departamento de
Física, Instituto Superior Técnico -- IST, Universidade de Lisboa -- UL, Avenida Rovisco Pais 1, 1049-001 Lisboa, Portugal}

\pacs{04.20.--q, 04.20.Jb, 04.40.Nr, 04.70.Bw}

\begin{abstract}
After a brief summary of the basic properties of stationary spacetimes representing rotating, charged black holes in strong axisymmetric magnetic fields, we concentrate on extremal cases, for which the horizon surface gravity vanishes. We investigate their properties by constructing simpler spacetimes that exhibit their geometries near degenerate horizons. Starting from the symmetry arguments we find that the near-horizon geometries of extremal magnetised Kerr-Newman black holes can be characterised by just one dimensionless parameter: \qt{effective Kerr-Newman mixing angle}. Employing the near-horizon geometries we demonstrate the Meissner effect of magnetic field expulsion from extremal black holes.
\end{abstract}

\maketitle
\tableofcontents

\section{Introduction and Conclusions}

Considerable attention has been paid to the interaction of black holes with external magnetic fields and charged particles from the mid-1970s already. This was motivated astrophysically after highly relativistic jets from active galactic nuclei and quasars were observed. There are various ways how a black hole or its accretion disc can power two opposite jets. The most relevant appears to be the Blandford-Znajek mechanism in which magnetic lines threading through the horizon of a rotating black hole are dragged by the black hole's rotation to spin around the rotation axis. As a consequence plasma is pushed outwards in opposite directions along the axis (see e.g. the review in \cite{membra}, more recently in \cite{Punsly, Penna1, Narayan}).

In the late 1960s and the 1970s the axially symmetric problem in the Einstein-Maxwell theory was reformulated and, using the symmetries of the field equations in the new formulation, generating techniques were invented. One of the most frequently applied is the so-called Harrison transformation \cite{Harr}. This led to the first papers by Ernst \cite{Ernst76, ErnstC} on spacetimes representing black holes in magnetic universes or accelerated charged black holes in an external electric field (cf. also \cite{ErnstWild, JiBi80}). In these spacetimes the gravitational back-reaction of the electromagnetic field is taken into account. 

The most general metric obtained by applying the Harrison transformation to the Kerr-Newman metric was written down by Bretón Baez and García Díaz in \cite{BaezDiaz86}. Therein the Harrison transformation is based on an arbitrary linear combination of the two Killing vectors, $\nicebpd{t}$ and $\nicebpd{\varphi}$, of the Kerr-Newman spacetime. The resulting metric involves a mixture of external electric and magnetic fields, and of electric and magnetic monopole charges on black hole. The final form is given by extremely lengthy expressions. We are not aware of any use of these metrics in the literature.

In the present work, as general magnetised black hole solutions the \qt{magnetised Kerr-Newman} (MKN) solutions will be analysed. They can be obtained by the Harrison transformation based on the axial Killing vector $\nicebpd{\varphi}$ (see Appendix \ref{app:ht}). These solutions coincide with those rederived recently by Gibbons, Mujtaba and Pope \cite{Pope} by the use of the $\mbb{SU}\(2,1\)$ global symmetry which arises after a Kaluza-Klein reduction of the four-dimensional Einstein-Maxwell theory. Our MKN solutions form a subclass in the branch $\mrm{MKN}\(\alpha=0,\beta=1\)$ of general solutions $\mrm{MKN}\(\alpha,\beta\)$ of \cite{BaezDiaz86}; here $\alpha$ and $\beta$ are coefficients in the linear combination of the Killing vectors $\nicebpd{t}$ and $\nicebpd{\varphi}$. It was established by Ernst \cite{Ernst76} that it is the use of the Harrison transformation based solely on the Killing vector $\nicebpd{\varphi}$ which leads to adding an external magnetic (and corresponding gravitational) field. By putting the magnetic charge $g=0$ in the branch $\mrm{MKN}\(\alpha=0,\beta=1\)$, one obtains solutions described by García Díaz in \cite{Diaz85}. To find these solutions García Díaz employs the Harrison transformation with a complex continuous parameter which implies the \qt{addition} of both electric and magnetic fields. Here we confine ourselves to pure magnetisation (cf. Appendix \ref{app:ht}). In this way we guarantee that the new solutions preserve the \qt{mirror symmetry}, i.e. are invariant under reflections $\vartheta\to\rpi-\vartheta$. For the same reason it is necessary to put the magnetic charge $p=0$ in the solutions of \cite{Pope}.

From the 1980s various aspects of rotating charged black holes with test or strong external fields were studied: the motion of test charged particles, the formation of an effective ergosphere, the behaviour of wave fields, etc. A comprehensive review of the results to which Gal'tsov and his colleagues contributed significantly can be found in \cite{AlijevGalcov}, where also many references are given. An important study of the structure of magnetised black holes was conducted by Hiscock \cite{Hiscock81}, who analysed the asymptotic behaviour of the dragging potential and electric field, concluding that \qt{the locally measured electric field on the symmetry axis in general approaches a constant, nonzero value far from the black hole.} 
There exists a simple solution of the Einstein-Maxwell equations, named after Melvin \cite{Melvin64}, which is static, cylindrically symmetric, with nontrivial magnetic but vanishing electric field. Hence, as Hiscock concludes \qt{the magnetized Kerr-Newman solution cannot globally (i.e., for all $\vartheta$) approach MMU}, i.e. the  Melvin magnetic universe. In \cite{Hiscock81} it was also demonstrated that \qt{the Schwarzschild-Melvin solution is unique among the Harrison-transformed magnetized Kerr-Newman black hole solutions in being static with no naked singularities.}

Most recently, magnetised black hole spacetimes were analysed in depth in the work of Gibbons, Mujtaba and Pope \cite{Pope} and Gibbons, Pang and Pope \cite{GibbonsPope2}. In the first paper, the authors note that \qt{the default assumption in the literature has been that this [Harrison-type generating techniques] will produce a background at infinity that is asymptotically Melvin}... That this is not necessarily so was anticipated in Hiscock's work \cite{Hiscock81} mentioned above. In \cite{Pope} a deeper, more detailed analysis of these aspects is given. For example, ergoregions in MKN spacetimes  are shown to extend from the black hole to infinity. Only for a specific relation among black the hole's charge, original seed value of the angular momentum and the value of the appended magnetic field the metric is asymptotic to the static Melvin metric \cite{Melvin64}. In \cite{GibbonsPope2} the thermodynamics of magnetised Kerr-Newman black holes is studied, in particular the formalism of how to calculate global quantities like mass and the angular momentum, which are suitable to formulate the first law of thermodynamics. The authors also show, by using the expressions for the electromagnetic field in the magnetised Kerr-Newman solution derived in \cite{Pope}, that if the physical charge on the black hole vanishes, the magnetic flux through the upper hemisphere of the horizon vanishes for extremal black holes. This result appears already in \cite{KarVok91}, where it is also shown that the magnetic flux vanishes when the angular momentum calculated from the Komar expression at the horizon vanishes.

This is the so-called black hole \qt{Meissner effect}: black holes approaching an extremal state expel external vacuum axially symmetric stationary (electro)magnetic fields. The effect is seen in the simplest situation with the magnetic test field, characterised by parameter $B$, which is uniform at infinity aligned with the rotation axis of a Kerr black hole \cite{Wald74}. The flux across the upper hemisphere of the horizon is equal to $\msc F_\mrm{H}=B\rpi r_+^2\(1-\nicefrac{a^4}{r_+^4}\)$, where $r_+=M+\sqrt{M^2-a^2}$ is the location of the outer horizon in the Boyer-Lindquist coordinates -- see \cite{King, BiJa85} -- hence, it vanishes when the black hole becomes extremal, i.e. $r_+=M=a$.
In fact, one can demonstrate that the black-hole Meissner effect is quite general: it takes place for \emph{all} axisymmetric stationary test fields around a rotating black hole \cite{BiJa85}; it arises also for extremal charged (non-rotating) black holes when even test (electromagnetic) fields are in general coupled to gravitational perturbations \cite{BiDvo80}. The expulsion also occurs in the exact models when external fields influence the spacetime geometry as shown around 1990 in \cite{BiKa, KarVok91}, later in \cite{KarBud}, and most recently in 2014 in \cite{GibbonsPope2} as mentioned above. The Meissner effect was also demonstrated for some extremal black hole solutions in higher dimensions in string theory and Kaluza-Klein theory \cite{ChEG}. A brief review until 2006 is contained in \cite{BiKaLe}. 

An important question of great astrophysical interest arises whether the Meissner effect decreases the efficiency of the extraction of energy from rotating black holes. The standard model of formation of highly relativistic jets from black holes requires magnetic field lines threading through the horizon. Due to the Meissner effect the fields are expelled from (nearly) extremal black holes, so the formation of jets could be quenched in this case. There are, however, field configurations which penetrate even into extremal black holes (non-axisymmetric fields \cite{BiJa85}, the split monopole fields \cite{Penna2}). Moreover, recent axisymmetric simulations of rotating black holes surrounded by magnetised plasma based on general-relativistic magnetohydrodynamics \qt{see no sign of expulsion of flux from the horizon} \cite{McK} although, as the authors mention, it is possible that they \qt{have not gone close enough to $\nicefrac{a}{M}=1$}. For the most recent review, summarising the simulations by various groups, see \cite{Narayan}. None of these simulations was performed for the extremal case. Takamori \emph{et al.} \cite{TNIKY} analysed the Meissner effect perturbatively for a stationary magnetosphere in an extremal black-hole background and found that if force-free electric current exists, higher multipole components can penetrate the extremal horizon. However, in their sophisticated analytic investigation the black hole was, in fact, an extremely charged \emph{static} black hole and the imposed test field was not treated consistently by using the coupled Einstein-Maxwell system as in \cite{BiDvo80}.

Regardless of its astrophysical relevance, the black-hole Meissner effect is a fundamental phenomenon the cause of which should be elucidated. In the case of test magnetic fields, the physics underlying the effect was discussed most recently in the works of Penna \cite{Penna2, Penna3}, from two different perspectives. In the first, the effect is related to the fact that the proper distance to the extremal horizon blows up and, in order to maintain the $\vec\nabla\cdot\vec B=0$ equation, the external field must vanish at such horizon (see Fig.~3 in \cite{Penna2}). In the second, the effect is derived from the low temperature limit of two-point correlation functions in the Hartle-Hawking vacuum, which imply that modes on either side of an extremal horizon become unentangled \cite{Penna3}.
In the following, we analyse the Meissner effect within the full Einstein-Maxwell theory in an alternative way -- by employing the near-horizon description of exact solutions representing extremal black holes in external magnetic fields characterised by parameter $B$.

The issue of describing the near-horizon geometry of extremal black holes has a long history. Indeed, some indications can already be found in the well-known work by Carter \cite{Carter68b}, in which metrics enabling a separable wave equation are derived. The Kerr metric is the best known example. However, Carter also includes (among different cases labelled by $\[\mrm{A}\], \[\mrm{B}\(-\)\]$, etc.) metrics which, in fact, represent near-horizon geometries. The transition between the cases is discussed formally only, without a physical interpretation.
Various formulations of near-horizon limit are possible with distinct interpretations related, for example, to the limiting behaviour of the Hawking temperature (see e.g. \cite{MalMichSt, Halilsoy93}). In our work, we \emph{start} with extremal MKN black holes with fixed physical parameters and use an arbitrary limiting parameter following the work by Bardeen and Horowitz \cite{BardHorow} on the Kerr-Newman spacetime. We, however, start from spacetimes which are not included in Carter's framework. That the near-horizon limit can be used beyond Kerr-Newman black holes has also been shown by Dias and Lemos in the case of accelerated black holes \cite{DiaLem03}. 

It is known that limiting metrics describing the near-horizon \qt{throat} regions usually have $\mrm{AdS}$-like asymptotics. This property makes the near-horizon limit interesting in string theory and holographic duality (see \cite{BardHorow, MalMichSt} and the living review by Compère \cite{Comp}). The high symmetry of the near-horizon limiting spacetimes has been analysed by Kunduri and Lucietti \cite{KuLuRe, KuLu13}, emphasising that the picture is similar even beyond $\mrm{4D}$ general relativity. In Appendix \ref{app:kt} we will give some explicit calculations that relate the well-known symmetry group $\mbb{SO}\f(1,2\)\times\mbb{U}\f(1\)$ generated by Killing vectors to the \qt{Carter-type} symmetry studied in \cite{Carter68b}. The fact that this kind of symmetry emerges in the near-horizon limit even when it is not present in the original spacetime will be useful to support our conclusions below. 

We found that there are some missing terms in the expressions for electromagnetic potential in Carter's fundamental work \cite{Carter68b} (in subcases $\[\mrm{B}\(\pm\)\]$). (This can also be seen by comparison with formulae for the $B^0_\pm$ subcases restricted to $f=1$ in \cite{Debever}.) Krasiński noticed the problem when he was editing Carter’s later work \cite{Carter73} for its republication \cite{Carter09} in the “Golden Oldie” series in the GRG journal. However, he did not relate the error to its root in the earlier article \cite{Carter68b}. Krasiński, in his editorial note \cite{editorialCarter}, interpreted Carter’s derivation as a de facto near-horizon limit, but he did not find the correct remedy for the error (whereas Bardeen and Horowitz \cite{BardHorow} did not discuss the behavior of the electromagnetic potential in the limit). We hope to give more comments on this
issue in a future work.
In the present work, in order to clarify these uncertainties, we rederive the process of the near-horizon limit step by step in Section \ref{sek:nhlim}. The general scheme was summarised by Compère \cite{Comp}.

In this paper we do not aim to study all theoretical implications of the near-horizon limit nor use it to examine intrinsic thermodynamic properties of the magnetised black holes. Instead, we employ it to investigate the interaction of the black holes with the external field in the strong field regime. We use the term near-horizon (limiting) \qt{description} in our work.

The outline of this article is as follows. In Section \ref{sek:uv} we briefly summarise some features of black hole solutions in magnetic universes, including the three cases admitting degenerate horizons, going from simpler stationary Ernst solution \cite{Ernst76} describing a Reissner-Nordström black hole in an external magnetic field, to the magnetised Kerr metric (i.e. Ernst-Wild solution \cite{ErnstWild}) and the general MKN black holes. We also mention possible gauges and the corresponding regularity of the electromagnetic potential at the axis -- which appears to be unnoticed in the literature so far.
We proceed to the near-horizon geometries of extremal cases in Section \ref{sek:nhlim}, using the general prescription and clarifying its details. 

It is known that the near-horizon geometry has four Killing vectors. In Appendix \ref{app:kt} we use them to construct a Killing tensor which is an element of symmetry related to the separability of the Hamilton-Jacobi (and the wave) equation. Since all the metrics admitting such symmetry were already derived by Carter, we conclude that there has to be some degeneracy and that it has to be possible to express the \qt{new} near-horizon geometries that we obtained (including the special case previously given in \cite{DiazBaez86}) using some simpler metrics with less parameters. We discuss this point in Section \ref{sek:nhdeg} going from special cases that are easy to express to the general extremal case of MKN black holes. 
We find that the near-horizon metric of extremal MKN black holes coincides (up to rescaling of Killing vectors by a constant) with the near-horizon Kerr-Newman metric described using just two independent effective parameters ($\hat M,\hat a,\hat Q$ minus the constraint of extremality) instead of three ($B, M, a, Q$ minus the constraint of extremality); the parameter $B$ characterising the strength of the external magnetic field enters expressions for $\hat a$ and $\hat Q$.
We can further reduce the number of parameters by excluding the physical scale and using dimensionless parameters. Then we end up with a graph of a plane with two parameters, $BM$ and $\gamma_\mrm{KN}$, where $\gamma_\mrm{KN}$ is the \qt{Kerr-Newman mixing angle}: $a=M\cos\gamma_\mrm{KN}, Q=M\sin\gamma_\mrm{KN}$. The plane is foliated by curves (classes of equivalence) labeled by just one parameter -- the \qt{effective Kerr-Newman mixing angle} $\hat\gamma_\mrm{KN}$. We note that this parameter is related to the invariants like the curvature of the horizon. 

We were led by the symmetry arguments given in Appendix \ref{app:kt} to conclude that the near-horizon description of any extremal MKN black hole is given by the near-horizon description of a corresponding extremal Kerr-Newman solution. However, this conclusion is also implied by the results of Lewandowski and Pawlowski \cite{LP}, who used the theory of isolated horizons to prove that all the extremal axially symmetric electrovacuum horizons must be the Kerr-Newman ones (for generalisations of this statement in the framework of near-horizon description, see the living review by Kunduri and Lucietti \cite{KuLu13}). In Section \ref{odd:inv} we sketch how the approach of \cite{LP} can also be used to define the effective parameters. 

Curiously enough, our expression for the product $\hat a\hat M$ coincides precisely with the angular momentum of a MKN black hole derived from general principles in \cite{GibbonsPope2}, when we restrict it to extremal cases. Our effective mass, however, does not match the one proposed in \cite{GibbonsPope2}. Booth \emph{et al.} \cite{Booth} recently inquired into (dis)agreements of various procedures of defining the mass of MKN black holes. 
One can see that our parameter $\hat M$ coincides with \qt{isolated horizon mass} $M_\mrm{IH}$ discussed in \cite{Booth}, when we evaluate it in the extremal case. 

In Section \ref{sek:nhdeg} we also discuss the Meissner effect. The external magnetic field strength parameter $B$ gets absorbed in the effective Kerr-Newman parameters. Hence, the magnetic flux coming through the degenerate horizon can be expressed without including $B$. Such flux is caused just by the physical charge on the black hole and its angular momentum. The effect of the external magnetic field is just an \qt{implicit} one. 

The way to generate a MKN solution by the Harrison transformation of the Kerr-Newman metric is summarised in Appendix \ref{app:ht}. Here it is also demonstrated that the rigidity theorems for dragging and electromagnetic potentials (and some related properties) are preserved by the transformation.\footnote{The paper is based on some results obtained in \cite{dipl} and during the Ph.D. study of F.H. under the supervision of J.B. A preliminary summary appeared in \cite{WDS}.}

\section{Charged, rotating black holes in magnetic fields}

\lbl{sek:uv}

As mentioned in the Introduction, a simple solution to the Einstein-Maxwell system of equations, often referred to as a \qt{magnetic universe}, was studied by Melvin \cite{Melvin64} (see also an earlier derivation in \cite{Bonnor}). The metric can be expressed in cylindrical coordinates ($R=r\sin\vartheta$, $z=r\cos\vartheta$) as
\rov{\mbs g=\(1+\frac{1}{4}B^2R^2\)^2\(-\bd t^2+\bd R^2+\bd z^2\)+\frac{R^2}{\(1+\frac{1}{4}B^2R^2\)^2}\bd\varphi^2\rvt}{mmu}
It is accompanied with the magnetic field, characterised by parameter $B$, with only non-zero component of the electromagnetic potential
\rov{A_\varphi=\frac{\frac{1}{2}BR^2}{1+\frac{1}{4}B^2R^2}\rvt}{afimmu}

The Harrison transformation applied on a \qt{seed} Minkowski spacetime yields this Melvin magnetic universe. If the Harrison transformation is applied to asymptotically flat black hole solutions, the results are black holes immersed in magnetic universes. These were studied by Ernst \cite{Ernst76}. 

\subsection{Stationary Ernst solution}

The stationary Ernst solution,
\rov{\mbs g=\left|\iLambda\right|^2\[-\(1-\frac{2M}{r}+\frac{Q^2}{r^2}\)\bd t^2+\frac{1}{1-\frac{2M}{r}+\frac{Q^2}{r^2}}\bd r^2+r^2\bd\vartheta^2\]+\frac{r^2\sin^2\vartheta}{\left|\iLambda\right|^2}\(\bd\varphi-\omega\bd t\)^2\rvc}{syE}
 represents a Reissner-Nordström black hole with mass parameter $M$ and charge parameter $Q$ in an external magnetic field. The influence of the magnetic field of a strength $B$ on the geometry is 
 expressed by a complex function $\iLambda$:
\rov{\iLambda=1+\frac{1}{4}B^2\(r^2\sin^2\vartheta+Q^2\cos^2\vartheta\)-\mrm iBQ\cos\vartheta\rvt}{lamsyE}
The electromagnetic field is more complicated than just a superposition of the electrostatic field of the black hole and some simple external field. This arises from the non-linear nature of the Einstein-Maxwell system. However, the actual field has the same crucial property as that simple superposition would have: a non-zero angular momentum. This induces frame dragging with dragging potential
\rov{\omega=-\frac{2BQ}{r}+B^3Qr+\frac{B^3Q^3}{2r}-\frac{B^3Q}{2r}\(r^2-2Mr+Q^2\)\sin^2\vartheta\rvt}{omsyE}

One can get the azimuthal component of the electromagnetic potential by means of the Harrison transformation (see, e.g. \cite{Pope})
\rov{A_\varphi^{\(1\)}=\frac{1}{B\left|\iLambda\right|^2}\[2\Re\iLambda\(\Re\iLambda-1\)+\(\Im\iLambda\)^2\]\rvt}{}
However, the given gauge is not very convenient because $A_\varphi^{\(1\)}$ does not vanish on the axis. We can fix this problem by subtracting the value of \qt{raw} $A_\varphi^{\(1\)}$ for $\vartheta=0$. In this way we obtain
\rov{A_\varphi=\frac{1}{B\left|\iLambda\right|^2}\[\frac{\(2+\frac{3}{2}B^2Q^2\)\(\Re\iLambda\)^2+\(1-\frac{1}{16}B^4Q^4\)\(\Im\iLambda\)^2}{1+\frac{3}{2}B^2Q^2+\frac{1}{16}B^4Q^4}-2\Re\iLambda\]\rvt}{AfisyE}
This unique gauge is important for the study of the particle motion within the Hamiltonian formalism, since there the electromagnetic potential becomes a physically relevant quantity. For the importance of having a regular electromagnetic potential at the axis, see also \cite{Comp}.

Another important quantity is the tetrad component describing the radial electric field strength
\rov{F_{(r)(t)}=\frac{1}{\left|\iLambda\right|^4}\left\{\[\(\Re\iLambda\)^2-\(\Im\iLambda\)^2\]\(2-\Re\iLambda\)\frac{Q}{r^2}+\frac{B}{2}\(1-\frac{Q^2}{r^2}\)\Im\iLambda\[\(\Re\iLambda\)^2-\(\Im\iLambda\)^2-4\Re\iLambda\]\cos\vartheta\right\}\rvt}{F20syE}
This directly enters some relations that we will discuss below.

\subsection{Ernst-Wild solution and a general MKN black hole}

In \cite{ErnstWild} the solution is obtained by applying the Harrison transformation to the Kerr (uncharged) \qt{seed} metric with rotation parameter $a$. The resulting metric 
\rov{\mbs{g}=\left|\iLambda\right|^2\iSigma\[-\frac{\iDelta}{\msc{A}}\bd t^2+\frac{\bd r^2}{\iDelta}+\bd\vartheta^2\]+\frac{\msc A}{\iSigma\left|\iLambda\right|^2}\sin^2\vartheta\(\bd\varphi-\omega\bd t\)^2}{ewmkn}
contains functions
\prov{\iSigma&=r^2+a^2\cos^2\vartheta\rvc&\msc A&=\(r^2+a^2\)^2-\iDelta a^2\sin^2\vartheta\rvc&\iDelta&=r^2-2Mr+a^2}
from the \qt{seed} metric and a new (complex) function 
\rov{\iLambda=1+\frac{1}{4}B^2\frac{\msc A}{\iSigma}\sin^2\vartheta-\frac{\mrm{i}}{2}B^2Ma\cos\vartheta\(3-\cos^2\vartheta+\frac{a^2}{\iSigma}\sin^4\vartheta\)\rvt}{lamew}
The new dragging potential is
\drov{\omega={}&\frac{a}{r^2+a^2}\Bigg\{\(1-B^4M^2a^2\)-\iDelta\Bigg[\frac{\iSigma}{\msc A}+\frac{B^4}{16}\Bigg(-8Mr\cos^2\vartheta\(3-\cos^2\vartheta\)-6Mr\sin^4\vartheta+\zrov&+\frac{2Ma^2\sin^6\vartheta}{\msc A}\[r\(r^2+a^2\)+2Ma^2\]+\frac{4M^2a^2\cos^2\vartheta}{\msc A}\[\(r^2+a^2\)\(3-\cos^2\vartheta\)^2-4a^2\sin^2\vartheta\]\Bigg)\Bigg]\Bigg\}\rvt}{omew}
The components of the field strength tensor are quite complicated. However, the azimuthal component of the potential has a very simple form in the gauge implied directly by the Harrison transformation:
\rov{A_\varphi^{(1)}=\frac{2}{B}\(1-\frac{\Re\iLambda}{\left|\iLambda\right|^2}\)\rvt}{}
After subtracting its value at the axis to obtain a regular expression we have
\rov{A_\varphi=\frac{2}{B}\(\frac{1}{1+B^4M^2a^2}-\frac{\Re\iLambda}{\left|\iLambda\right|^2}\)\rvt}{}

The general MKN (magnetised Kerr-Newman) black hole has the metric \eqref{ewmkn} with $\iDelta$ from the Kerr-Newman metric, i.e. $\iDelta=r^2-2Mr+a^2+Q^2$, and with a more complicated function $\iLambda$:
\drov{\iLambda=1&+\frac{1}{4}B^2\(\frac{\msc A+a^2Q^2\(1+\cos^2\vartheta\)}{\iSigma}\sin^2\vartheta+Q^2\cos^2\vartheta\)+\zrov&+\frac{BQ}{\iSigma}\[ar\sin^2\vartheta-\mrm{i}\(r^2+a^2\)\cos\vartheta\]-\frac{\mrm{i}}{2}B^2a\cos\vartheta\[M\(3-\cos^2\vartheta\)+\frac{Ma^2\sin^2\vartheta-Q^2r}{\iSigma}\sin^2\vartheta\]\rvt}{lammkn}
The most complete and comprehensive picture of the properties of a general MKN black hole was recently given by Gibbons, Mujtaba and Pope \cite{Pope}. The expression for the dragging potential is given by equations (B.8)-(B.9) in \cite{Pope}, whereas the components of electromagnetic potential are described by formulae (B.15)-(B.18).

\subsection{Global properties of magnetised black holes}

First, let us note that one can easily verify that the positions of the Killing horizons in the coordinate $r$ are left in place by the Harrison transformation. They are still given by the roots of $\iDelta$, i.e. $r_\pm=M+\sqrt{M^2-Q^2-a^2}$. The roots coincide for $M^2=Q^2+a^2$, which defines the extremal case; one can make sure that the surface gravity of the Killing horizon vanishes in this case. These facts were already noted e.g. in \cite{KarVok91}.

The asymptotic properties of the magnetised black-hole spacetimes can be complicated for large values of dimensionless quantity $BM$; in that case an astrophysical meaning is doubtful. The conditions for the existence of an approximately flat region have been discussed by Bičák and Janiš \cite{BiJa85}. The region must be well outside the horizon, but it must also hold that $\left|\iLambda\right|^2$ is approximately unity in that region. These two requirements are satisfied when $r$ satisfies inequality $r_+\ll r\ll \nicefrac{1}{B}$. This is well visualised by the embedding diagrams constructed in \cite{StuchHle99}. 

As first noted by Hiscock \cite{Hiscock81}, a conical singularity along the axis of symmetry arises when the Harrison transformation is applied to the metrics of the Kerr-Newman family. In order to remove it, we have to adjust the range for the azimuthal coordinate to $\varphi\in\[0,\varphi_\mrm{max}\)$, where
\rov{\varphi_\mrm{max}=2\rpi\lim_{\vartheta\to0}\frac{\sqrt{g_{\vartheta\vartheta}\f(r,\vartheta\)}}{\pder{\sqrt{g_{\varphi\varphi}\f(r,\vartheta\)}}{\vartheta}}\rvt}{csrem}
In case of a general MKN black hole we get
\rov{\varphi_\mrm{max}=2\rpi\[1+\frac{3}{2}B^2Q^2+2B^3MQa+B^4\(\frac{1}{16}Q^4+M^2a^2\)\]\rvt}{azrange}
Alternatively, we can pass to the rescaled azimuthal coordinate $\tilde\varphi=\nicefrac{\(2\rpi\varphi\)}{\varphi_\mrm{max}}$, which runs in the standard range $(0,2\rpi)$. However, we shall follow most of the literature and use $\varphi$ with \eqref{azrange}. 

\section{The near-horizon description of extremal configurations}

\lbl{sek:nhlim}

We now generalise the approach of \cite{Carter73, BardHorow, Comp} to find the near-horizon geometries of extremal black holes and apply it to black holes in strong magnetic fields. In the case of the MKN black holes we impose the extremality condition $M^2=Q^2+a^2$.

\subsection{General prescription}

\lbl{odd:nhrec}

\subsubsection{The metric in the extremal case}

The metric of an axisymmetric, stationary, electrovacuum black hole can be written as follows \cite{FrolNov}:
\rov{\mbs g=-N^2\bd t^2+g_{\varphi\varphi}\(\bd\varphi-\omega\bd t\)^2+g_{rr}\bd r^2+g_{\vartheta\vartheta}\bd\vartheta^2\rvt}{axst}
For an extremal black hole the metric can be cast into the form
\rov{\mbs g=-\(r-r_0\)^2\tilde N^2\bd t^2+g_{\varphi\varphi}\(\bd\varphi-\omega\bd t\)^2+\frac{\tilde g_{rr}}{\(r-r_0\)^2}\bd r^2+g_{\vartheta\vartheta}\bd\vartheta^2\rvc}{axst2}
where the degenerate horizon is located at $r=r_0$ and $\tilde N$ and $\tilde g_{rr}$ are regular and non-vanishing at the horizon.

To describe the \qt{near-horizon} region, we introduce new time and spatial coordinates $\tau$ and $\chi$ by relations
\prov{r&=r_0+p\chi\rvc&t&=\frac{\tau}{p}\rvt\label{rchittau}}
The transformation depends on a limiting parameter $p$; for any finite non-zero value of $p$ the new coordinates cover the entire spacetime up to \qt{standard}  spatial infinity. However, in the extremal black-hole spacetimes, there exists yet another infinity: the proper radial distance between two points along $t=\mrm{constant}$ diverges if one of the points approaches $r_0$. With parameter $p$ in transformation \eqref{rchittau} converging to zero, the metric \eqref{axst2} goes over to a new metric which describes the infinite region (\qt{throat}) involving $r=r_0$. The standard spatial infinity is \qt{lost} in this  limiting procedure.

A more complicated issue arises in describing the dragging in the near-horizon limit. To do this we first expand $\omega$ around its value $\omega_\mrm{H}$ at the horizon 
\rov{\omega\doteq\omega_\mrm{H}+\res{\pder{\omega}{r}}{r_0}\(r-r_0\)=\omega_\mrm{H}+\res{\pder{\omega}{r}}{r_0}p\chi\rvc}{omnhexp}
so that
\rov{\bd\varphi-\omega\bd t\doteq\bd\varphi-\(\omega_\mrm{H}+\res{\pder{\omega}{r}}{r_0}p\chi\)\frac{\bd\tau}{p}=\bd\varphi-\frac{\omega_\mrm{H}}{p}\bd\tau-\res{\pder{\omega}{r}}{r_0}\chi\bd\tau\rvt}{dragnhexp}
Since as a consequence of the rigidity theorem (see Appendix \ref{app:ht}) the value $\omega_\mrm{H}$ is constant,
the transformation from $\varphi$ to an \qt{unwinded angle} $\psi$ can be integrated
\rov{\varphi=\psi+\frac{\omega_\mrm{H}}{p}\tau\rvt}{phipsi}
With $p\to0$, we obtain
\rov{\bd\varphi-\omega\bd t=\bd\psi-\res{\pder{\omega}{r}}{r_0}\chi\bd\tau\rvt}{dragnhexp2}
It is seen that the new dragging potential, after the limit $p\to0$, is just the first order of the expansion of the original $\omega$. 
Note that \eqref{phipsi} is merely a transformation to (rigidly) rotating coordinates: it can be understood as a \qt{gauge-fixing} of the integration constant of the dragging potential. 

Following this procedure, we arrive at the near-horizon metric given in Appendix \ref{app:kt}, equation \eqref{nhsp}. In the specific case of the Kerr-Newman solution, the resulting metric reads
\rov{\mbs g=\[Q^2+a^2\(1+\cos^2\vartheta\)\]\(-\frac{\chi^2}{\(Q^2+2a^2\)^2}\bd\tau^2+\frac{\bd\chi^2}{\chi^2}+\bd\vartheta^2\)+\frac{\(Q^2+2a^2\)^2\sin^2\vartheta}{Q^2+a^2\(1+\cos^2\vartheta\)}\(\bd\psi+\frac{2a\sqrt{Q^2+a^2}\chi}{\(Q^2+2a^2\)^2}\bd\tau\)^2\rvc}{knnh}
(cf. \cite{Carter73} and, in particular, \cite{BardHorow}, where the equation above with small rearrangements is formula (4.2)).

\subsubsection{The electromagnetic field}

As a consequence of the limiting procedure with $p\to0$, we cannot expect that every tensor quantity will be regular after this limit. In some cases such problems can be circumvented as it happens with the electromagnetic potential. Defining the generalised electrostatic potential by
\rov{\phi=-A_t-\omega A_\varphi\rvc}{elstgen}
and expanding the electromagnetic potential in $r-r_0=p\chi$ using equations \eqref{rchittau}, \eqref{phipsi}, we obtain
\drov{\mbs A=A_t\bd t+A_\varphi\bd\varphi&\doteq\(\res{A_t}{r_0}+\res{\pder{A_t}{r}}{r_0}p\chi\)\frac{\bd\tau}{p}+\(\res{A_\varphi}{r_0}+\res{\pder{A_\varphi}{r}}{r_0}p\chi\)\(\bd\psi+\frac{\omega_\mrm{H}}{p}\bd\tau\)\doteq\zrov&\doteq\(\res{A_t}{r_0}+\omega_\mrm{H}\res{A_\varphi}{r_0}\)\frac{\bd\tau}{p}+\(\res{\pder{A_t}{r}}{r_0}+\omega_\mrm{H}\res{\pder{A_\varphi}{r}}{r_0}\)\chi\bd\tau+\res{A_\varphi}{r_0}\bd\psi=\zrov&=-\frac{\phi_\mrm{H}}{p}\bd\tau+\(\res{\pder{A_t}{r}}{r_0}+\omega_\mrm{H}\res{\pder{A_\varphi}{r}}{r_0}\)\chi\bd\tau+\res{A_\varphi}{r_0}\bd\psi\rvt}{emnhlim}
Since $\phi_\mrm{H}=\mrm{constant}$ (see Appendix \ref{app:ht}), the singular first term can be subtracted from the final limiting potential as a gauge constant. 

For the particular case of the Kerr-Newman solution the near-horizon electromagnetic potential (without the singular gauge constant) reads\footnote{Bardeen and Horowitz \cite{BardHorow} do not give the expression for the electromagnetic potential. Carter's potential (see equation (5.64) in the original version in \cite{Carter73}) does not satisfy Maxwell equations as noted in the \qt{Editorial Note} \cite{editorialCarter}. Unfortunately, the potential given in the republication  \cite{Carter09} of \cite{Carter73} is also not correct. The correct form of the potential, though given in an other rearrangement, is contained in equation (39) in \cite{Comp}.} 
\rov{\mbs A=\frac{Q}{Q^2+a^2\(1+\cos^2\vartheta\)}\(\frac{Q^2+a^2\sin^2\vartheta}{Q^2+2a^2}\chi\bd\tau+a\sqrt{Q^2+a^2}\sin^2\vartheta\bd\psi\)\rvt}{aknnh}

Let us turn back to the case of axially symmetric, stationary, electrovacuum black hole with metric \eqref{axst} and electromagnetic potential \eqref{emnhlim}. Introducing the frame vectors associated with metric \eqref{axst}
\prov{\mbs e_{(t)}&=\frac{1}{N}\(\bpd{t}+\omega\bpd{\varphi}\)\rvc&\mbs e_{(\varphi)}&=\frac{1}{\sqrt{g_{\varphi\varphi}}}\bpd{\varphi}\rvc\lbl{lnrf1}\\\mbs e_{(r)}&=\frac{1}{\sqrt{g_{rr}}}\bpd{r}\rvc&\mbs e_{(\vartheta)}&=\frac{1}{\sqrt{g_{\vartheta\vartheta}}}\bpd{\vartheta}\rvc\lbl{lnrf2}}
we find that the frame component of the radial electric field strength reads
\rov{F_{(r)(t)}=\frac{1}{N\sqrt{g_{rr}}}\(\pder{A_t}{r}+\omega\pder{A_\varphi}{r}\)\rvt}{}
This component can be obtained directly from the Harrison transformation just by algebraic manipulation and derivatives (cf. equation \eqref{htF20}). Regarding \eqref{emnhlim}, we can express the time component of the electromagnetic potential after $p\to0$ as
\rov{A_\tau=\res{\(\tilde N\sqrt{\tilde g_{rr}}F_{(r)(t)}\)}{r_0}\chi\rvt}{ataunh1}

Now it is important to realise that on a degenerate horizon of any extremal MKN black hole (see Appendix \ref{app:ht})
\prov{\tilde\omega&\equiv\res{\pder{\omega}{r}}{r=r_0}=\mrm{constant}\rvc&%\textrm{and}&&
\tilde\phi&\equiv\res{\pder{\phi}{r}}{r=r_0}=\mrm{constant}\rvt\lbl{nhtiqu0}}

Using equations \eqref{dragnhexp2}, \eqref{nhtiqu0} and realizing that the expression multiplying $\chi$ in equation \eqref{ataunh1} is just a function of $\vartheta$, we can write in the near-horizon limit
\prov{\omega&=\tilde\omega\chi\rvc&\phi&=\tilde\phi\chi\rvc&A_\tau&=\tilde A_\tau\f(\vartheta\)\chi\rvt\lbl{nhtiqu}}
From the definition of the generalised electrostatic potential \eqref{elstgen} we can express
\prov{\phi&=-A_\tau-\omega A_\psi\rvc&A_\psi\f(\vartheta\)&=\frac{1}{\tilde\omega}\(-\tilde\phi-\tilde A_\tau\f(\vartheta\)\)\rvc}
where, observing \eqref{emnhlim}, we put $A_\psi=\res{A_\varphi}{r_0}$. Since we demand the electromagnetic potential to be smooth, its azimuthal component must vanish at the axis, so
\prov{\phi&=-\res{A_\tau}{\vartheta=0}\rvc&A_\psi\f(\vartheta\)&=\frac{1}{\tilde\omega}\(\tilde A_\tau\f(0\)-\tilde A_\tau\f(\vartheta\)\)\rvt\lbl{apsi1}}

We conclude that the knowledge of $F_{(r)(t)}$ in the original spacetime is sufficient to determine $A_\tau$ (see \eqref{ataunh1}) and hence the whole potential of electromagnetic field in the near-horizon limit by relations \eqref{apsi1}. 

\subsection{Near-horizon description of extremal black holes in strong magnetic fields}

We shall now apply the above procedure to black holes in strong magnetic fields going from the simplest stationary Ernst solution to the MKN black holes. 

\subsubsection{Stationary Ernst solution}

\lbl{odd:nhsye}

Applying the recipe described above to metric \eqref{syE}, we get
\drov{\mbs g=\[\(1+\frac{1}{4}B^2Q^2\)^2+B^2Q^2\cos^2\vartheta\]\(-\frac{\chi^2}{Q^2}\bd\tau^2+\frac{Q^2}{\chi^2}\bd\chi^2+Q^2\bd\vartheta^2\)+\zrov+\frac{Q^2\sin^2\vartheta}{\(1+\frac{1}{4}B^2Q^2\)^2+B^2Q^2\cos^2\vartheta}\[\bd\psi-\frac{2B}{Q}\(1+\frac{1}{4}B^2Q^2\)\chi\bd\tau\]^2\rvc}{syEnh}
which for $B=0$ turns to \eqref{knnh} with $a=0$, i.e. to the Robinson-Bertotti solution. The components of the electromagnetic potential can be evaluated directly from $F_{(r)(t)}$ given in \eqref{F20syE}:
\prov{A_\tau&=\frac{\chi}{Q}\frac{\(1+\frac{1}{4}B^2Q^2\)^2-B^2Q^2\cos^2\vartheta}{\(1+\frac{1}{4}B^2Q^2\)^2+B^2Q^2\cos^2\vartheta}\(1-\frac{1}{4}B^2Q^2\)\rvc\lbl{AtausyEnh}\\A_\psi&=\frac{-BQ^2}{1+\frac{3}{2}B^2Q^2+\frac{1}{16}B^4Q^4}\frac{\(1-\frac{1}{16}B^4Q^4\)\sin^2\vartheta}{\(1+\frac{1}{4}B^2Q^2\)^2+B^2Q^2\cos^2\vartheta}\rvt\lbl{ApsisyEnh}}
We checked explicitly that these quantities satisfy the full Einstein-Maxwell system.\footnote{It should be noted that this simple solution is contained in a richer class of solutions studied in \cite{DiazBaez86}. There the generalisations of Reinssner-Nordström and Robinson-Bertotti solutions employ Harrison transformation involving more parameters. Our expressions \eqref{syEnh}-\eqref{ApsisyEnh} can be obtained from equations (4.1)-(4.5) in \cite{DiazBaez86}, if we choose $a=1$, $b=h=k=0$, $\alpha=0$, $\beta=1$, rescale the metric by $M^2$; physical parametres in \cite{DiazBaez86} $g=B=0$ and $e$ and $E$ are equal to our $\nicefrac{Q}{M}$ and $\nicefrac{-BM}{2}$; coordinates $\tau, q, p$, and $\sigma$, respectively, are put equal to our $\nicefrac{\tau}{M},\nicefrac{\chi}{M},\cos\vartheta$, and $\psi$, respectively. At the end of \cite{DiazBaez86}, a near horizon limit of \qt{magnetic Reissner-Nordström} solution is shown to imply a \qt{magnetised Robinson-Bertotti solution}. The near-horizon limit of the Ernst-Wild and MKN solutions analysed in the following is not contained in \cite{DiazBaez86}.}

\subsubsection{Ernst-Wild solution}

Applying the near-horizon limit to metric \eqref{ewmkn}, one gets
\drov{\mbs g=\[\(1+B^4a^4\)\(1+\cos^2\vartheta\)+2B^2a^2\sin^2\vartheta\]\(-\frac{\chi^2}{4a^2}\bd\tau^2+\frac{a^2}{\chi^2}\bd\chi^2+a^2\bd\vartheta^2\)+\zrov+\frac{4a^2\sin^2\vartheta}{\(1+B^4a^4\)\(1+\cos^2\vartheta\)+2B^2a^2\sin^2\vartheta}\[\bd\psi+\frac{\chi}{2a^2}\(1-B^4a^4\)\bd\tau\]^2\rvt}{ewnh}
Note that the dragging potential \eqref{omew} vastly simplifies in the near-horizon limit. From the $F_{(r)(t)}$ component of the field strength tensor (see equation \eqref{htF20}) one obtains 
\prov{A_\tau&=-B\chi\(1-\frac{2\(1+B^2a^2\)^2}{\(1+B^4a^4\)\(1+\cos^2\vartheta\)+2B^2a^2\sin^2\vartheta}\)\rvc\\A_\psi&=\frac{1-B^4a^4}{1+B^4a^4}\frac{2Ba^2\sin^2\vartheta}{\(1+B^4a^4\)\(1+\cos^2\vartheta\)+2B^2a^2\sin^2\vartheta}\rvt}
Again, one can check that quantities evaluated above satisfy the full Einstein-Maxwell system. 

\subsubsection{A general MKN black hole}

Although the Ernst-Wild solution is more complicated than the stationary Ernst solution, we have seen that their near-horizon descriptions are likewise simple. Turning finally to the general MKN case, we find that such simplicity is lost. The metric can be written in the form (cf. equation \eqref{mknnhgen} in Appendix \ref{app:kt})
\rov{\mbs g=\vld f\f(\vartheta\)\(-\frac{\chi^2}{\(Q^2+2a^2\)^2}\bd\tau^2+\frac{\bd\chi^2}{\chi^2}+\bd\vartheta^2\)+\frac{\(Q^2+2a^2\)^2\sin^2\vartheta}{\vld f\f(\vartheta\)}\(\bd\psi-\tilde\omega\chi\bd\tau\)^2\rvc}{mknnhdim}
where (after rearrangements enabling to make \qt{division} by denominator $Q^2+a^2\(1+\cos^2\vartheta\)$) the dimensional structural function $\vld f\f(\vartheta\)$ is obtained as
\drov{\vld f\f(\vartheta\)={}&\(1+\frac{1}{4}B^2Q^2\)^2\[Q^2+a^2\(1+\cos^2\vartheta\)\]+2\(1+\frac{1}{4}B^2Q^2\)\[B^2a^2\(Q^2+a^2\)+BQa\sqrt{Q^2+a^2}\]\sin^2\vartheta+\zrov&+B^2\(Ba\sqrt{Q^2+a^2}+Q\)^2\[a^2+\(Q^2+a^2\)\cos^2\vartheta\]\rvt}{mknstrdim0}
This can still be simplified: 
\rov{\vld f\f(\vartheta\)=\[\sqrt{Q^2+a^2}\(1+\frac{1}{4}B^2Q^2+B^2a^2\)+BQa\]^2+\[a\(1-\frac{3}{4}B^2Q^2-B^2a^2\)-BQ\sqrt{Q^2+a^2}\]^2\cos^2\vartheta\rvt}{mknstrdim}
The dragging constant $\tilde\omega$ can be evaluated using equation \eqref{htomr} in the following form:
\rov{\tilde\omega=-\frac{2}{\(Q^2+2a^2\)^2}\[\sqrt{Q^2+a^2}\(1+\frac{1}{4}B^2Q^2+B^2a^2\)+BQa\]\[a\(1-\frac{3}{4}B^2Q^2-B^2a^2\)-BQ\sqrt{Q^2+a^2}\]\rvt}{tommkn}
From relations \eqref{htF20} and \eqref{ataunh1} for the $F_{(r)(t)}$ component we obtain:
\prov{A_\tau={}&\frac{1}{\vld f\(\vartheta\)}\frac{\chi}{Q^2+2a^2}\[Q\(1-\frac{1}{4}B^2Q^2\)+2Ba\sqrt{Q^2+a^2}\]\vast\{\Bigg[Q\(1-\frac{1}{4}B^2Q^2\)+\lbl{ataumknnh}\zrovn&+2Ba\sqrt{Q^2+a^2}\Bigg]^2+\Bigg[a\(1-\frac{3}{4}B^2Q^2-B^2a^2\)-BQ\sqrt{Q^2+a^2}\Bigg]^2\sin^2\vartheta\vast\}\rvc\\A_\psi={}&-\frac{\tilde\omega}{2\vld f\f(\vartheta\)}\frac{(Q^2+2a^2)^2\[Q\(1-\frac{1}{4}B^2Q^2\)+2Ba\sqrt{Q^2+a^2}\]\sin^2\vartheta}{1+\frac{3}{2}B^2Q^2+2B^3Qa\sqrt{Q^2+a^2}+B^4\(\frac{1}{16}Q^4+Q^2a^2+a^4\)}\lbl{apsimknnh}\rvt}
Notice that even in this most complicated case, $A_\psi$ contains the overall $\sin^2\vartheta$ factor. 

\section{Near-horizon degeneracy of extremal MKN class}

\lbl{sek:nhdeg}

\subsection{Special cases}

\lbl{odd:crossdeg}

Starting from the previous section, we first investigate important special cases. First, notice that in the near-horizon description of the Ernst-Wild solution \eqref{ewnh}, the dragging parameter $\tilde\omega=0$ for $Ba=1$. The metric becomes static and the factor $\[\(1+B^4a^4\)\(1+\cos^2\vartheta\)+2B^2a^2\sin^2\vartheta\]=4$, so the metric simplifies significantly. After some rearrangements we find
\rov{\mbs g=-\frac{\chi^2}{\(2a\)^2}\(2\bd\tau\)^2+\frac{\(2a\)^2}{\chi^2}\bd\chi^2+\(2a\)^2\bd\vartheta^2+\(2a\)^2\sin^2\vartheta\(\frac{\bd\psi}{2}\)^2\rvt}{EWRBrang}
It is noteworthy that the electromagnetic field turns out to be purely electric in this case since the azimuthal component of the potential (given by \eqref{AtausyEnh} and \eqref{ApsisyEnh}) vanishes:
\rov{\mbs A=\frac{\chi}{2a}\(2\bd\tau\)+0\frac{\bd\psi}{2}\rvt}{}
This special case is thus the Robinson-Bertotti spacetime with an electric constant $2a$, which is expressed in the rescaled time and azimuth. (The rescaling of azimuth does not lead to a conical singularity because $\psi$ has the same range as  $\varphi$ from the original Ernst-Wild solution; this runs from $0$ to $4\rpi$ due to \eqref{azrange}.)

The fact that the limiting near-horizon metric is static and spherically symmetric has \emph{local} implications for the original spacetime -- it is easy to see that the embedding of the cross section of the horizon ($r=M,t=\mrm{constant}$) of the Ernst-Wild solution with $a=M, Ba=1$ is indeed a two-sphere. 

If we choose $Ba=-1$, we get the same results as above with $A_\mu$  exchanged by $-A_\mu$. 

The structure of the near-horizon metric \eqref{mknnhgen} is simple. It involves a single function $f\f(\vartheta\)$ and two constants $K$ and $\tilde\omega$. The Robinson-Bertotti solution with $f\equiv1$, $\tilde\omega=0$ and arbitrary $K$ is still the simpler case of this class. Another such case arises when $2f\f(\vartheta\)=1+\cos^2\vartheta$ and $\tilde\omega=\nicefrac{-1}{K^2}$ (the minus sign means rotation in the sense of positive $\psi$). This case is the near-horizon description of an extremal Kerr solution, as can be seen from \eqref{knnh} with $Q$ set to zero. In the canonical form \eqref{mknnhgen} it reads 
\rov{\mbs g=\frac{1+\cos^2\vartheta}{2}\(-\frac{\chi^2}{2a^2}\bd\tau^2+\frac{2a^2}{\chi^2}\bd\chi^2+2a^2\bd\vartheta^2\)+\frac{2a^2}{\frac{1}{2}\(1+\cos^2\vartheta\)}\sin^2\vartheta\(\bd\psi+\frac{\chi}{2a^2}\bd\tau\)^2\rvt}{knh}

Remarkably, we arrive at the same metric starting from the special case of a stationary Ernst solution. Regarding potential \eqref{AtausyEnh} and \eqref{ApsisyEnh} one easily sees that the potential vanishes if we choose parameters $B$ and $Q$ such that $BQ=2$. Function $f\f(\vartheta\)$ simplifies to
\rov{\res{f\f(\vartheta\)=\[\(1+\frac{1}{4}B^2Q^2\)^2+B^2Q^2\cos^2\vartheta\]}{B=\frac{2}{Q}}=4\(1+\cos^2\vartheta\)\rvc}{}
and metric \eqref{syEnh}, after rearrangements, becomes
\rov{\mbs g=\frac{1+\cos^2\vartheta}{2}\[-\frac{\chi^2}{2\(2Q\)^2}\(8\bd\tau\)^2+\frac{2\(2Q\)^2}{\chi^2}\bd\chi^2+2\(2Q\)^2\bd\vartheta^2\]+\frac{2\(2Q\)^2\sin^2\vartheta}{\frac{1}{2}\(1+\cos^2\vartheta\)}\[\frac{\bd\psi}{8}+\frac{\chi}{2\(2Q\)^2}\(8\bd\tau\)\]^2\rvt}{syEknrang}
Hence, the near-horizon metric for the stationary Ernst solution with $B=\nicefrac{2}{Q}$ turns to the Kerr near-horizon form \eqref{knh} with rotation parameter $2Q$. Notice that the condition $\tilde\omega=\nicefrac{-1}{K^2}$ emerges only in appropriately rescaled coordinates; this is related to the range of $\psi$ preserving smooth axis. 

In Section \ref{odd:bezroz} we shall see that even a general MKN class admits special combinations of parameters $Q,a,B$ for which either the dragging parameter $\tilde\omega$ or electromagnetic potential vanish in the near-horizon limit.
However, already from the simple examples above, we conclude that \qt{bare} parameters $Q, a, B$ do not describe properly physical properties of MKN black holes in the near-horizon limit. Since we do not observe any magnetic field in the near-horizon description in the special cases, we can be led to an idea that external axially symmetric stationary magnetic fields are expelled from degenerate horizons. 

\subsection{Effective parameters}

\lbl{odd:eff}

It is known that near-horizon limiting spacetimes possess high symmetry exhibited by four Killing vectors (cf. e.g. \cite{Comp}). In Appendix \ref{app:kt} we show explicitly that it is possible to construct a \qt{Carter-type} Killing tensor from these Killing vectors and therefore the near-horizon spacetimes belong to the class of spacetimes studied by Carter \cite{Carter68b, Carter73}, even in the case when the original spacetime is outside this class. Therefore, it is natural to assume that the metrics \eqref{knnh} and \eqref{mknnhdim} representing near-horizon limits of extremal Kerr-Newman and MKN black holes are mathematically equivalent in general. It should be possible to describe the near-horizon limit of extremal black holes in magnetic fields characterised by three parameters ($M, a, Q, B$ minus the constraint of extremality) by just two effective \qt{Kerr-Newman-like} parameters ($\hat M,\hat a,\hat Q$ minus the constraint of extremality). 
However, metrics \eqref{knnh} and \eqref{mknnhdim} cannot be compared directly, since their Killing vectors are scaled in a different way, as we observed in special cases already.

Now we shall rescale the coordinates in general MKN cases. It is convenient to redefine $\varphi_\mrm{max}$ in \eqref{azrange} by introducing $\iXi=\nicefrac{2\rpi}{\varphi_\mrm{max}}$, so that $\psi\in\(0,\nicefrac{2\rpi}{\iXi}\)$. In the extremal case  
\rov{\iXi=\frac{1}{1+\frac{3}{2}B^2Q^2+2B^3Qa\sqrt{Q^2+a^2}+B^4\(\frac{1}{16}Q^4+Q^2a^2+a^4\)}\rvt}{azrange2}
Next we rearrange the constants in metric \eqref{mknnhdim} in analogy with \eqref{EWRBrang} and \eqref{syEknrang} to obtain
\rov{\mbs g=\vld f\f(\vartheta\)\[-\frac{\iXi^2\chi^2}{K^4}\(\frac{\bd\tau}{\iXi}\)^2+\frac{\bd\chi^2}{\chi^2}+\bd\vartheta^2\]+\frac{K^4\sin^2\vartheta}{\iXi^2\vld f\f(\vartheta\)}\[\(\iXi\bd\psi\)-\iXi^2\tilde\omega\chi\frac{\bd\tau}{\iXi}\]^2\rvc}{mknnhrang2}
where \qt{new} azimuthal coordinate $\iXi\psi\in\(0,2\rpi\)$ and, correspondingly, the \qt{new} time coordinate becomes $\nicefrac{\tau}{\iXi}$. Under such rescaling the components of the electromagnetic potential become
\rov{\mbs A=\iXi A_\tau\frac{\bd\tau}{\iXi}+\frac{A_\psi}{\iXi}\(\iXi\bd\psi\)\rvt}{amknnhrang}

Since the rescaling leaves function $\vld f\f(\vartheta\)$ in \eqref{mknstrdim} unchanged, we can identify the coefficients appearing inside function $\vld f\f(\vartheta\)$ in accordance with the limiting near-horizon metric of the Kerr-Newman solution \eqref{knnh} which is $\vld f\f(\vartheta\)=\[Q^2+a^2\(1+\cos^2\vartheta\)\]=M^2+a^2\cos^2\vartheta$. Hence, we are able to write $\vld f\f(\vartheta\)$ in \eqref{mknstrdim} as 
\rov{\vld f\f(\vartheta\)={\hat M}^2+{\hat a}^2\cos^2\vartheta\rvc}{maeffdef}
where the resulting effective parameters $\hat M$ and $\hat a$ are
\prov{\hat M&=\sqrt{Q^2+a^2}\(1+\frac{1}{4}B^2Q^2+B^2a^2\)+BQa\rvc\lbl{meff}\\\hat a&=a\(1-\frac{3}{4}B^2Q^2-B^2a^2\)-BQ\sqrt{Q^2+a^2}\rvt\lbl{aeff}}
In analogy with  the Kerr-Newman solution, we define the effective charge by ${\hat Q}^2={\hat M}^2-{\hat a}^2$, which implies 
\rov{\hat Q=Q\(1-\frac{1}{4}B^2Q^2\)+2Ba\sqrt{Q^2+a^2}\rvt}{qeff}
It will be seen that $\hat Q$ is a physical charge.

To demonstrate that the near-horizon descriptions of MKN black holes and of corresponding Kerr-Newman black holes are indeed equivalent, we replace $Q$ and $a$ in \eqref{knnh} by expressions for $\hat Q, \hat a$ \eqref{qeff} and \eqref{aeff}. We find that the result is the limiting MKN metric \eqref{mknnhdim} with coordinates rescaled according to the formula \eqref{mknnhrang2}. Then we do the same for the electromagnetic potential, i.e. insert the expressions for $\hat Q, \hat a$ in place of $Q, a$ in \eqref{aknnh} and compare with \eqref{ataumknnh} and \eqref{apsimknnh}, again with coordinates rescaled the same way (see \eqref{amknnhrang}). Such a procedure requires lengthy calculations but using Mathematica we were able to verify that the description of the near-horizon limit of a MKN black hole using the effective parameters indeed works. The solution \eqref{mknnhdim} is thus mathematically equivalent to the solution \eqref{knnh} with effective parameters $\hat M, \hat a, \hat Q$. 

Regarding the expressions \eqref{meff}-\eqref{qeff} for the effective parameters, it is of interest to notice that each of them contains the \qt{bare} parameter multiplied by a \qt{correction} involving $B^2$ and an additional term linear in $B$. The extra term in the rotation parameter \eqref{aeff} is caused by the angular momentum of the electromagnetic field as can be seen from putting the bare rotation parameter $a=0$. The extra term in charge \eqref{qeff} corresponds to the \qt{Wald charge} in the weak-field limit (cf. \cite{Wald74}). 

The effective parameters simplify considerably if one of the bare parameters $Q$ or $a$ is zero -- they involve just one term. This is reflected in the fact that the near-horizon descriptions of the stationary Ernst solution \eqref{syEnh} and of the Ernst-Wild solution \eqref{ewnh} are much simpler than of a general MKN solution. For $B=0$, the effective parameters are equal to the bare parameters. It is also worthwhile to notice that when one solves the equation $\hat M=0$ with respect to $B$, one finds that it has no real roots, so $\hat M$ cannot be negative. 

Concerning the effective charge, it can be shown that it is the physical charge integrated over the horizon. Indeed, Karas and Vokrouhlický \cite{KarVok91} discuss various integral quantities for a MKN black hole, including its physical charge
\rov{Q_\mrm{H}=\frac{1}{4\rpi}\int\limits_0^{\rpi}\int\limits_0^{\varphi_\mrm{max}}\star F_{\vartheta\varphi}\td\varphi\td\vartheta\rvt}{}
Using their formula (9) and evaluating the Ernst potentials (cf. Appendix \ref{app:ht}) for a \emph{general} MKN black hole we obtain
\rov{Q_\mrm{H}=Q\(1-\frac{1}{4}B^2Q^2\)+2BMa\rvc}{qphys}
where $M\geqslant\sqrt{Q^2+a^2}$ (cf. also \cite{Pope}). In the extremal case, $M=\sqrt{Q^2+a^2}$, and it is seen explicitly from \eqref{qeff} that $\hat Q=Q_\mrm{H}$.

It is remarkable that when we define the angular momentum of extremal MKN black holes by $\hat J=\hat a\hat M$, i.e. in analogy with the standard case without magnetic field, we find that the result coincides precisely with the thermodynamic angular momentum given by Gibbons, Pang and Pope \cite{GibbonsPope2} in formula (5.11), when it is restricted to the extremal case.\footnote{Let us note that the expression $-2\hat J$ appeared already in the curly bracket in formula (2.47) in thesis \cite{dipl}, which shortly preceded \cite{GibbonsPope2}. However, we did not even hint at the interpretation of $\hat J$ as a meaningful angular momentum.} We already know that, in properly rescaled coordinates, the near-horizon limit of any extremal MKN black hole can be described just by effective parameters $\hat Q, \hat a$. We can substitute for $\hat a$ using $\hat Q,\hat J$ as follows:
\rov{\hat a=\frac{\sgn\hat J}{\sqrt{2}}\sqrt{\sqrt{{\hat Q}^4+4{\hat J}^2}-{\hat Q}^2}\rvt}{}
Since $\hat Q$ is a physical charge of the black hole and $\hat J$ is derived in \cite{GibbonsPope2} as a meaningful angular momentum, we can conclude that the near-horizon limit of any extremal MKN black hole can be described by thermodynamic charges of the black hole. 

However, a puzzle remains: whereas our \qt{near-horizon mass} $\hat M=\sqrt{{\hat Q}^2+{\hat a}^2}$, the thermodynamic mass evaluated in \cite{GibbonsPope2} by means of Kaluza-Klein reduction is found to be $\nicefrac{M}{\iXi}$ (with $\iXi$ given by \eqref{azrange2} in the extemal case) and, therefore, is a distinct parameter. 
The problem of (dis)agreements among different notions of mass in the MKN spacetimes has been recently discussed by Booth \emph{et al.} \cite{Booth}. By a closer inspection of their results, one can make sure that our $\hat M$ coincides with the \qt{isolated horizon mass} $M_\mrm{IH}$ defined in \cite{Booth}, when we restrict it to the extremal case\footnote{In \cite{Booth} the authors claim that their $M_\mrm{IH}$ is equal to the Komar mass of the MKN horizon. The Komar integrals for the MKN black holes were discussed also in \cite{KarVok91} long time ago. The results disagree. This is apparently caused by the fact that \cite{Booth} uses different, more sophisticated forms of the Komar expressions.}.
 
\subsection{Dimensionless parameters}

\lbl{odd:bezroz}

As a consequence of the extremality condition, one of three (bare) parameters $M, a, Q$ is not independent.\footnote{The same holds for the corresponding effective parameters.} 
Instead of choosing two of these as independent parameters, it is more convenient to introduce the \qt{Kerr-Newman mixing angle} $\gamma_\mrm{KN}$ by relations 
\prov{a&=M\cos\gamma_\mrm{KN}\rvc&Q=&M\sin\gamma_\mrm{KN}\rvt\lbl{exknmix}}
Here $\gamma_\mrm{KN}$ can be taken from the interval $\[-\nicefrac{\rpi}{2},\nicefrac{\rpi}{2}\]$, if we restrict the rotation parameter to $a>0$.
Since the mass is just a scale, it is ignorable. As another dimensionless parameter we consider $BM$.
These two parameters, $\gamma_\mrm{KN}$ and $BM$, represent dimensionless quantities that define the parameter space. We now proceed to analyse its structure.

In Section \ref{odd:crossdeg} we considered a special case of the near-horizon description of Ernst-Wild solution for which the dragging parameter $\tilde\omega=0$. During our proof of the equivalence of the near-horizon descriptions of a general extremal MKN black hole and the corresponding Kerr-Newman black hole (see the text below \eqref{qeff}) we found $\tilde\omega\sim\hat M\hat a$. From this it is evident that the case with $\tilde\omega=0$ within general extremal MKN black holes can be obtained just by putting $\hat a=0$ in \eqref{aeff}. We find two branches of the solution with respect to $B$ as follows:
\rov{B=\frac{-2Q\sqrt{Q^2+a^2}\pm2\(Q^2+2a^2\)}{a\(3Q^2+4a^2\)}\rvt}{brbeq}
From equation \eqref{maeffdef} it follows immediately that these two subclasses have the near-horizon geometry of the Robinson-Bertotti spacetime. Moreover, equations \eqref{ataumknnh} and \eqref{apsimknnh} indeed imply that $A_\psi=0$ and the component $A_\tau$ yields constant electric field. In our dimensionless parameters the solutions \eqref{brbeq} read
\rov{BM=\frac{-2\sin\gamma_\mrm{KN}\pm2\(1+\cos^2\gamma_\mrm{KN}\)}{3\cos\gamma_\mrm{KN}+\cos^3\gamma_\mrm{KN}}\rvt}{brbeq1}
Let us remark that in \cite{KarVok91} Komar angular momentum for an extremal MKN black hole is evaluated; using our notation we see that it vanishes for $\tilde\omega=0$. Our result \eqref{brbeq} coincides with (12) in \cite{KarVok91}. However, therein the results were derived differently -- formula \eqref{brbeq} was obtained by requiring the vanishing magnetic flux accross an \qt{upper hemisphere} of a horizon. We shall return to this Meissner-type effect below. 

Let us turn to another special subclass of the MKN solutions. Since both components of the electromagnetic potential, \eqref{ataumknnh} and \eqref{apsimknnh}, for the MKN solutions in the near-horizon limit are proportional to  the parameter $\hat Q$ given in \eqref{qeff}, we can make them vanish by putting $\hat Q=0$. 
This condition implies
\rov{B=\frac{4a\sqrt{Q^2+a^2}\mp2\(Q^2+2a^2\)}{Q^3}\rvt}{bvaceq}
(The $\mp$ sign corresponds to $\tilde\omega=\nicefrac{\mp1}{\iXi K^2}$.) In terms of dimensionless parameters we get
\rov{BM=\frac{4\cos\gamma_\mrm{KN}\mp2\(1+\cos^2\gamma_\mrm{KN}\)}{\sin^3\gamma_\mrm{KN}}\rvt}{bvaceq1}
Notice that in this case $\hat Q=0$ implies the physical charge \eqref{qphys} also vanishes. In fact, as demonstrated in \cite{KarVok91}, the physical charge $Q_\mrm{H}=0$ even for a non-extremal MKN black hole if $B=2Q^{-3}\(2Ma\pm\sqrt{4M^2a^2+Q^4}\)$.
In the extremal case, this implies our result \eqref{bvaceq}. 

Regarding the parametrisation \eqref{exknmix}, we can view the special subclasses as curves in the parameter space of $\gamma_\mrm{KN}$ and $BM$ given by formulae \eqref{brbeq1} and \eqref{bvaceq1} respectively. Moreover, we know that using the effective parameters $\hat M, \hat a,\hat Q$ (see \eqref{meff}-\eqref{qeff}) the near-horizon description of \emph{any} extremal MKN black hole can be expressed as near-horizon description of the corresponding extremal Kerr-Newman solution (see the text below \eqref{qeff}). Therefore the entire parameter space can be foliated by curves which represent equivalent near-horizon geometries. This is illustrated in Figure \ref{MKNps}. Four of such curves, representing special subclasses ($\hat a=0$ and $\hat Q=0$) analysed above, are given by four special values of dimensionless parameter $\hat\gamma_\mrm{KN}$ introduced in analogy with \eqref{exknmix} by 
\rov{\hat Q=\hat M\sin\hat\gamma_\mrm{KN}\rvt}{mknfol0}
Substituting for $\hat Q$ and $\hat M$ from \eqref{qeff} and \eqref{meff} and using the previous relation to express $B$ we obtain 
\rov{B=\frac{4a\sqrt{Q^2+a^2}-2Qa\sin\hat\gamma_\mrm{KN}\mp2\(Q^2+2a^2\)\cos\hat\gamma_\mrm{KN}}{Q^3+\(Q^2+4a^2\)\sqrt{Q^2+a^2}\sin\hat\gamma_\mrm{KN}}\rvt}{}
Equivalently, in terms of dimensionless $\gamma_\mrm{KN}$ and $BM$,
\rov{BM=\frac{4\cos\gamma_\mrm{KN}-\sin2\gamma_\mrm{KN}\sin\hat\gamma_\mrm{KN}\mp2\(1+\cos^2\gamma_\mrm{KN}\)\cos\hat\gamma_\mrm{KN}}{\sin^3\gamma_\mrm{KN}+\(1+3\cos^2\gamma_\mrm{KN}\)\sin\hat\gamma_\mrm{KN}}\rvt}{}
The $\mp$ sign is not necessary if we extend the interval for $\hat\gamma_\mrm{KN}$ to  $\(-\rpi,\rpi\]$. Then the minus sign in front of the term with $\cos\hat\gamma_\mrm{KN}$ gurantees that $\hat\gamma_\mrm{KN}=0$ implies $\hat a=\hat M$. 

We can write equation \eqref{mknfol0} in an alternative form using $\hat a$ and $\hat M$ as $\hat a=\hat M\cos\hat\gamma_\mrm{KN}$. Then we obtain
\rov{B=\frac{-2Q\sqrt{Q^2+a^2}-2Qa\cos\hat\gamma_\mrm{KN}\pm2\(Q^2+2a^2\)\sin\hat\gamma_\mrm{KN}}{a\(3Q^2+4a^2\)+\(Q^2+4a^2\)\sqrt{Q^2+a^2}\cos\hat\gamma_\mrm{KN}}\rvc}{}
or 
\rov{BM=\frac{-2\sin\gamma_\mrm{KN}-\sin2\gamma_\mrm{KN}\cos\hat\gamma_\mrm{KN}\pm2\(1+\cos^2\gamma_\mrm{KN}\)\sin\hat\gamma_\mrm{KN}}{3\cos\gamma_\mrm{KN}+\cos^3\gamma_\mrm{KN}+\(1+3\cos^2\gamma_\mrm{KN}\)\cos\hat\gamma_\mrm{KN}}\rvt}{}

\begin{figure}
\centering
\input{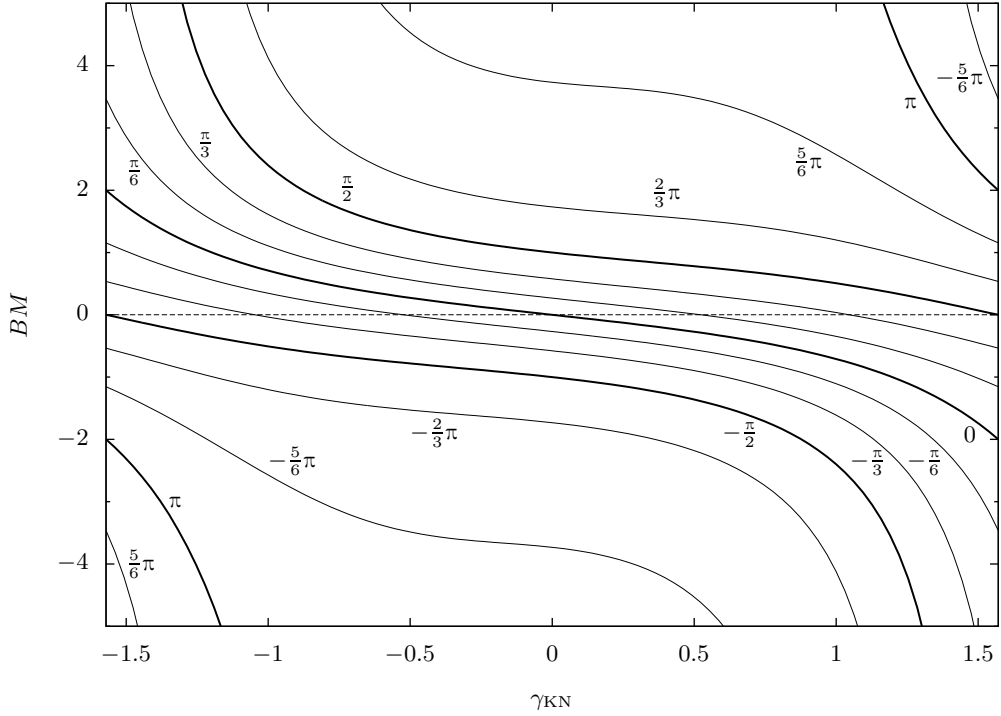}
\caption{Near-horizon geometries of extremal MKN black holes illustrated in the dimensionless parameter space $\(\gamma_\mrm{KN},BM\)$. Each curve describes geometries with fixed parameter $\hat\gamma_\mrm{KN}$, i.e. with fixed ratios $\nicefrac{\hat Q}{\hat M}$ and $\nicefrac{\hat a}{\hat M}$ (see equation \eqref{mknfol0}).Various regions in the plane correspond to different signs of parameters $\hat a,\hat Q$, respectively. The boundaries between the regions are indicated by thick lines.}
\lbl{MKNps}
\end{figure}

Karas and Vokrouhlický \cite{KarVok91} found that in the two special cases with special relations among parameters $B, Q, a$ given by expressions \eqref{brbeq} and \eqref{bvaceq} the magnetic flux vanishes corresponding to the black hole Meissner effect discovered in the test-field limit. In these cases our results employing the near-horizon description support the conclusion given in \cite{KarVok91}, since the magnetic field is encoded in the component $A_\psi$ of the electromagnetic potential, which is proportional to the product $\hat Q\hat a$. The two cases \eqref{brbeq} and \eqref{bvaceq} correspond to $\hat a=0$ and $\hat Q=0$, respectively. Hence $A_\psi=0$ and magnetic field necessarily vanishes. 

For a general extremal MKN black hole the magnetic flux through the upper hemisphere of the horizon does not vanish. It can be expressed as follows:
\rov{\msc F_\mrm{H}=2\rpi\frac{\res{A_\psi}{\vartheta=\frac{\rpi}{2}}}{\iXi}=2\rpi\frac{\hat Q\hat a}{\hat M}=\frac{4\rpi\hat Q\hat J}{{\hat Q}^2+\sqrt{{\hat Q}^4+4{\hat J}^2}}\rvt}{}
Since the structure of the azimuthal component of the electromagnetic potential in near-horizon limit is identical (up to the rescaling) to the one of the Kerr-Newman black hole, the flux can be expressed using the Kerr-Newman-like effective parameters. As stated above, these parameters can be related to the thermodynamic charges of the black hole as derived in \cite{GibbonsPope2}. Therefore, we may conclude that the magnetic flux is intrinsic to the black hole configuration and there is no flux caused directly by the external magnetic field. 

\subsection{Remark on invariants and uniqueness theorems}

\lbl{odd:inv}

As we have seen, due to the ambiguity of the scaling of the Killing vectors in the MKN spacetime, we have to choose a particular rescaling in order to see that the near-horizon limit is given by a corresponding Kerr-Newman solution with effective parameters given above. However, invariants are unaffected by coordinate transformations of \emph{\qt{ignorable}} coordinates and, therefore, they can be in principle used to determine the effective parameters. For example, consider the invariant $R_\mrm{2D}=\nicefrac{2}{\rho_1\rho_2}$, where $\rho_1,\rho_2$ are curvature radii of the (axially symmetric) degenerate horizon, given by the two-dimensional Ricci scalar
\rov{R_\mrm{2D}=\res{\frac{2}{g_{\vartheta \vartheta }}\[\frac{g_{\vartheta \vartheta ,\vartheta }g_{\varphi\varphi,\vartheta }}{4g_{\vartheta \vartheta }g_{\varphi\varphi}}-\frac{g_{\varphi\varphi,\vartheta \vartheta }}{2g_{\varphi\varphi}}+\frac{\(g_{\varphi\varphi,\vartheta }\)^2}{4\(g_{\varphi\varphi}\)^2}\]}{r_0}\rvt}{}
In the near-horizon limit the angular metric functions have no spatial dependence, so
\rov{R_\mrm{2D}=\frac{2}{g_{\vartheta \vartheta }}\[\frac{g_{\vartheta \vartheta ,\vartheta }g_{\psi\psi,\vartheta }}{4g_{\vartheta \vartheta }g_{\psi\psi}}-\frac{g_{\psi\psi,\vartheta \vartheta }}{2g_{\psi\psi}}+\frac{\(g_{\psi\psi,\vartheta }\)^2}{4\(g_{\psi\psi}\)^2}\]\rvt}{}
It is explicitly seen that altering $g_{\psi\psi}$ by a multiplicative factor (due to the linear rescaling of the coordinate $\psi$) does not change the result. 

For Kerr-Newman solution the invariant turns out to be
\rov{\frac{2}{R_\mrm{2D}}=\frac{\[Q^2+a^2\(1+\cos^2\vartheta\)\]^3}{\(Q^2+2a^2\)\[Q^2+a^2\(1-3\cos^2\vartheta\)\]}\rvc}{}
and hence for a MKN solution the form will be the same with $\hat Q, \hat a$ in place of $Q, a$. 

Although these expressions are not uniquely invertible, they show that there exists a connection between the effective parameters and the curvature invariants.

We used symmetry arguments stated in Appendix \ref{app:kt} to support the idea that the near-horizon geometry of any extremal MKN black hole can be described using near-horizon geometry of a corresponding Kerr-Newman solution. We should, however, note that Lewandowski and Pawlowski \cite{LP} conducted a sophisticated discussion of electrovacuum extremal horizons including uniqueness theorem from which the mentioned statement arises as a special case.\footnote{Indeed, in the recent preprint, \cite{Booth} Booth \emph{et al.}, led by the considerations in \cite{LP}, give our formulae \eqref{meff} and \eqref{aeff} (which also appeared already in the first summary of our results in 2014 -- see \cite{WDS}). Note that in \cite{Booth} the effective parameters are defined by a formula analogous to our equation \eqref{maeffdef}, rather than in the way suggested here in Section \ref{odd:inv}.} 
Namely they solved the constraint equations for all axially symmetrical extremal isolated horizons equipped with axially symmetrical electromagnetic field invariant with respect to the null flow (i.e. the field that is also in equilibrium). Having set the magnetic charge to zero they found out that their family of solutions has just two parameters, one of which is the area of the horizon and the other encodes the electric charge (they did not separate out the physical scale as we did in our subcase with the definition of $\hat\gamma_\mrm{KN}$). Since there exists a Kerr-Newman extremal horizon for each pair of values of their parameters, Lewandowski and Pawlowski conclude that there are no other extremal isolated electrovacuum horizons than the Kerr-Newman ones (see also discussion of similar results including non-zero cosmological constant in the framework of near-horizon geometries given in \cite{KuLu13}). 

The area of the horizon $\msc S_\mrm{H}$ used in \cite{LP} to parametrise their solutions is an integral invariant. Unlike the example of a differential invariant that we examined above, it can be used to give an alternative definition of the effective parameters, 
 as we would like to sketch here. The area of the horizon af any MKN black hole is
\rov{\msc S_\mrm{H}=4\rpi\frac{r_+^2+a^2}{\iXi}\rvt}{}
Therefore the magnetic field influences it just via factor $\iXi$, which assures the smoothnes of the axis. In the extremal case $r_+^2+a^2=Q^2+2a^2$ and $\iXi$ is given by \eqref{azrange2}. From the fact that the area of the horizon of an extremal MKN black hole must be characterised by Kerr-Newman effective parameters, it follows the relation
\rov{\frac{Q^2+2a^2}{\iXi}={\hat Q}^2+2{\hat a}^2\rvc}{equivK}
which is also obtained when we directly compare the MKN near-horizon metric in the form \eqref{mknnhrang2} with the Kerr-Newman near-horizon metric in the form \eqref{mknnhgen} with effective parameters plugged in. If we assume that the effective charge is the physical charge enclosed in the horizon $\hat Q=Q_\mrm{H}$, we can use formula \eqref{equivK} as a defining relation for $\hat a$. Finally, we can define $\hat M=\sqrt{{\hat Q}^2+{\hat a}^2}$.

\section*{Acknowledgement}

We thank Pavel Krtouš and David Kofroň for helpful discussions. J.B. acknowledges the support from the Czech Science Foundation, grant No. GAČR 14-37086G (Albert Einstein Centre), F.H. was supported by grants of the Charles University in Prague No. GAUK 606412, No. SVV 260211 and No. SVV 265301. We also thank the Albert Einstein Institute in Golm for the hospitality during a brief but important period of our collaboration.

\appendix

\section{Killing vectors and tensors in the near-horizon spacetimes}

\lbl{app:kt}

The metric of a near-horizon spacetime obtained by the procedure described in Section \ref{odd:nhrec} has the form 
\rov{\mbs g=-\chi^2\res{\tilde N^2}{r_0}\f(\vartheta\)\bd\tau^2+\res{g_{\varphi\varphi}}{r_0}\f(\vartheta\)\(\bd\psi-\tilde\omega\chi\bd\tau\)^2+\frac{\res{\tilde g_{rr}}{r_0}\f(\vartheta\)}{\chi^2}\bd\chi^2+\res{g_{\vartheta\vartheta}}{r_0}\f(\vartheta\)\bd\vartheta^2\rvc}{nhgen}
where the metric \qt{functions} come from the original spacetime. In the models we analyse it is easy to see that 
\rov{\res{\frac{\tilde g_{rr}}{\tilde N^2}}{r_0}=K^4}{}
is independent of $\vartheta$. So the metric \eqref{nhgen} can always be written as follows:
\rov{\mbs g=f\f(\vartheta\)\(-\frac{\chi^2}{K^2}\bd\tau^2+\frac{K^2}{\chi^2}\bd\chi^2\)+\res{g_{\varphi\varphi}}{r_0}\f(\vartheta\)\(\bd\psi-\tilde\omega\chi\bd\tau\)^2+\res{g_{\vartheta\vartheta}}{r_0}\f(\vartheta\)\bd\vartheta^2\rvc}{nhsp}
where we choose the structural function $f\f(\vartheta\)$ non-negative. 

For the MKN class the near-horizon limiting metric can be expressed in even more special form (cf. \eqref{mknstrdim})
\rov{\mbs g=f\f(\vartheta\)\(-\frac{\chi^2}{K^2}\bd\tau^2+\frac{K^2}{\chi^2}\bd\chi^2+K^2\bd\vartheta^2\)+\frac{K^2\sin^2\vartheta}{f\f(\vartheta\)}\(\bd\psi-\tilde\omega\chi\bd\tau\)^2\rvt}{mknnhgen}

The near-horizon limit preserves stationarity and axial symmetry exhibited by Killing vectors $\mbs\xi_{(1)}=\nicebpd{\tau}$ and $\mbs\xi_{(4)}=\nicebpd{\psi}$.
Regarding the Killing equation, it turns out that the Killing vector of \qt{anti-de Sitter type} (cf. \cite{BardHorow})
\rov{\mbs\xi_{(2)}=\tau\bpd{\tau}-\chi\bpd{\chi}}{kvnh1}
is guaranteed to exist by \eqref{nhgen}, as well as another Killing vector of anti-de Sitter type,
\rov{\mbs\xi_{(3)}=\(\frac{K^4}{2\chi^2}+\frac{\tau^2}{2}\)\bpd{\tau}-\tau\chi\bpd{\chi}+\tilde\omega\frac{K^4}{\chi}\bpd{\psi}\rvc}{kvnh2}
which involves constant $K$. We conclude that all the Killing vectors described by Bardeen and Horowitz \cite{BardHorow} for the Kerr case exist in a general near-horizon spacetime (see also \cite{Comp}). We give the expressions \eqref{kvnh1},\eqref{kvnh2} explicitly here to use them to construct Killing tensors. 

The Killing vectors $\xi_{(1)}^\mu, \xi_{(2)}^\mu, \xi_{(3)}^\mu$ with algebra $\kom{\mbs\xi_{(1)}}{\mbs\xi_{(2)}}=\mbs\xi_{(1)}$, $\kom{\mbs\xi_{(1)}}{\mbs\xi_{(3)}}=\mbs\xi_{(2)}$, $\kom{\mbs\xi_{(2)}}{\mbs\xi_{(3)}}=\mbs\xi_{(3)}$
are generators of the $\mbb{SO}\f(1,2\)$ part of the symmetry group,  whereas the axial Killing vector $\xi_{(4)}^\mu$ generates the $\mbb{U}\f(1\)$ part. We have thus reproduced, within our class of solutions, the result of \cite{KuLuRe}.\footnote{Let us note that the results of \cite{KuLuRe} are generalised in a recent work \cite{KuLu13} which admits an interesting special case, when anti-de Sitter properties of the near-horizon spacetime turn to de Sitter ones with the structural function $f\f(\vartheta\)$ of \eqref{nhsp} being non-positive.} 
 
The enhancement of symmetry and the similarity to the Kerr case pose naturally the question, whether the \qt{Carter-type} constant of motion\footnote{We mean a constant of motion consistent with the stationarity of dynamics. Killing vectors $\xi_{(2)}^\mu$ and $\xi_{(3)}^\mu$ have time-dependent components and so the constants of motion induced by them do not conform with the stationarity.} yielding the complete separability of geodesic equations exists in the near-horizon spacetimes. We demonstrate that the separation can be achieved in detail elsewhere (see \cite{dipl} and the paper in preparation), here we just construct the Killing tensor which is related to the separation constant $\tilde{\msc K}$.

To construct it we start from the well-known fact that a symmetrised tensor product of two Killing vectors yields a Killing tensor. First step is to find a non-trivial Killing tensor with time-independent components from the Killing vectors available. The correct way to do so is to use all three Killing vectors of anti-de Sitter kind, which yields
\prov{\zeta_{(0)}^{\iota\kappa}&=\xi_{(1)}^\iota\xi_{(3)}^\kappa+\xi_{(3)}^\iota\xi_{(1)}^\kappa-\xi_{(2)}^\iota\xi_{(2)}^\kappa\rvc\\\zeta_{(0)}^{\alpha\beta}\bpd{x^\alpha}\bpd{x^\beta}&=\frac{K^4}{\chi^2}\(\bpd{\tau}\)^2-\chi^2\(\bpd{\chi}\)^2+\tilde\omega\frac{K^4}{\chi}\(\bpd{\tau}\bpd{\psi}+\bpd{\psi}\bpd{\tau}\)\rvt}
Covariant components of this Killing tensor can be simplified by adding the tensor product of the fourth Killing vector with itself: 
\prov{\zeta^{(1)}_{\mu\nu}&=\xi^{(1)}_\mu\xi^{(3)}_\nu+\xi^{(3)}_\mu\xi^{(1)}_\nu-\xi^{(2)}_\mu\xi^{(2)}_\nu+\tilde\omega^2K^4\xi^{(4)}_\mu\xi^{(4)}_\nu\rvc\\\zeta^{(1)}_{\kappa\lambda}\bd x^\kappa\bd x^\lambda&=f^2\f(\vartheta\)\(\chi^2\bd\tau^2-\frac{K^4}{\chi^2}\bd\chi^2\)\rvt}
Now we can take the last step by employing the normalisation of a test particle velocity. We arrive at the Killing tensor 
\rov{\zeta^{(2)}_{\mu\nu}=\xi^{(1)}_\mu\xi^{(3)}_\nu+\xi^{(3)}_\mu\xi^{(1)}_\nu-\xi^{(2)}_\mu\xi^{(2)}_\nu+\tilde\omega^2K^4\xi^{(4)}_\mu\xi^{(4)}_\nu+f_{(0)}K^2g_{\mu\nu}\rvc}{}
where we assumed the decomposition {$f\f(\vartheta\)=f_{(0)}+f_{(1)}\f(\vartheta\)$} with $f_{(0)}=\mrm{constant.}$ 
It holds
\rov{\zeta^{(2)}_{\alpha\beta}u^\alpha u^\beta=\tilde{\msc K}\rvt}{}

Regarding how the Carter constant in the Kerr-Newman spacetime is also induced by a Killing tensor (cf. \cite{Wald, Carter73}), we can make sure that the Killing tensor $\zeta_{\iota\kappa}^\mrm{C}$ which induces the Carter constant in the Kerr-Newman spacetime reduces, in the near-horizon limit, to the Killing tensor $\zeta^{(2)}_{\iota\kappa}$ defined above (see also \cite{GalOr}). 

\section{Harrison transformation}

\lbl{app:ht}

Here we give some relations between the quantities in the \qt{magnetised} spacetime generated by the Harrison transformation and quantities in the original, \qt{seed} spacetime. We also show how the rigidity theorems are preserved in the transformation from the Kerr-Newman solution to a general MKN black hole. For details on the Harrison transformation, see \cite{Harr, Ernst76, ErnstWild} and \cite{Stephani}.

\subsection{Generating \qt{magnetised} solutions}

The complex Ernst gravitational potential for the Kerr-Newman solution that we employ reads
\drov{\msc E=-\(r^2+a^2-a\frac{2Ma+\mrm{i}\(2Mr-Q^2\)\cos\vartheta}{r+\mrm{i}a\cos\vartheta}\)\sin^2\vartheta-\(4Ma^2+\mrm{i}Q^2\cos\vartheta\)\frac{a-\mrm{i}r\cos\vartheta}{r+\mrm{i}a\cos\vartheta}\rvc}{grernstkn}
 the corresponding complex Ernst electromagnetic potential is
\rov{\iPhi=Q\frac{a-\mrm{i}r\cos\vartheta}{r+\mrm{i}a\cos\vartheta}\rvt}{emernstkn}
The Harrison transformation that we use is based on the axial Killing vector $\nicebpd{\varphi}$ and employs a real continuous parameter $B$. Utilising the potentials $\msc E$, $\iPhi$ and parameter $B$ we construct a complex function $\iLambda$ by 
\rov{\iLambda=1+B\iPhi-\frac{1}{4}B^2\msc E\rvt}{lamgen}
The transformation consists in the transition from the potentials $\msc E, \iPhi$ to the new potentials $\msc E', \iPhi'$ representing a new solution. It has the following form:
\prov{\msc E'&=\frac{\msc E}{\iLambda}\rvc&\iPhi'&=\frac{\iPhi-\frac{1}{2}B\msc E}{\iLambda}\lbl{htphi}\rvt}

The new dragging potential $\omega'$ (again a real quantity) for a general magnetised solution is given by two real partial differential equations
\prov{\pder{\omega'}{r}&=\left|\iLambda\right|^2\pder{\omega}{r}+\frac{2\sqrt{-g}}{g_{\varphi\varphi}g_{\vartheta\vartheta}}\(\Re\iLambda\pder{\Im\iLambda}{\vartheta}-\pder{\Re\iLambda}{\vartheta}\Im\iLambda\)\rvc\lbl{htomr}\\
\pder{\omega'}{\vartheta}&=\left|\iLambda\right|^2\pder{\omega}{\vartheta}-\frac{2\sqrt{-g}}{g_{\varphi\varphi}g_{rr}}\(\Re\iLambda\pder{\Im\iLambda}{r}-\pder{\Re\iLambda}{r}\Im\iLambda\)=\zrov&=\left|\iLambda\right|^2\pder{\omega}{\vartheta}-2\frac{N\sqrt{g_{\vartheta\vartheta}}}{\sqrt{g_{\varphi\varphi}g_{rr}}}\(\Re\iLambda\pder{\Im\iLambda}{r}-\pder{\Re\iLambda}{r}\Im\iLambda\)\rvc\lbl{htomth}}
where quantities involving the metric coefficients are taken from the original (\qt{seed}) metric (cf. equation \eqref{axst}), $\Re$ and $\Im$ denote real and imaginary parts.

Real and imaginary parts of potential $\iPhi$ are related to the components of the electromagnetic field as follows:
\prov{\Re\iPhi&=A_\varphi\rvc&\pder{\Im\iPhi}{r}&=-\sqrt{g_{\varphi\varphi}g_{rr}}F_{(\vartheta)(t)}\rvc&\pder{\Im\iPhi}{\vartheta}=\sqrt{g_{\varphi\varphi}g_{\vartheta\vartheta}}F_{(r)(t)}\rvt\lbl{emernstgen}}
(This holds both for the original and transformed, \qt{primed} quantities. Note that under the Harrison transformation based on $\nicebpd{\varphi}$ the products $g_{\varphi\varphi}g_{rr}$ and $g_{\varphi\varphi}g_{\vartheta\vartheta}$ do not change.)
Here $A_\varphi$ is the (coordinate) azimuthal component of the electromagnetic potential, whereas $F_{(\vartheta)(t)}$ and $F_{(r)(t)}$ are frame components of the field strength tensor in the locally non-rotating tetrad \eqref{lnrf1}-\eqref{lnrf2}. We substitute 
 for $\iPhi'$ from \eqref{htphi}, with $\msc E, \iPhi$  and $\iLambda$ expressed from \eqref{grernstkn}, \eqref{emernstkn} and \eqref{lamgen}. Equation \eqref{emernstgen} then implies 
\rov{F_{(r)(t)}'=\frac{1}{\sqrt{g_{\varphi\varphi}g_{\vartheta\vartheta}}}\pder{}{\vartheta}\Bigg[\frac{1}{\left|\iLambda\right|^2}\(-\Im\iLambda\Re\iPhi+\frac{1}{2}B\Im\iLambda\Re\msc E+\Re\iLambda\Im\iPhi-\frac{\mrm{1}}{2}B\Re\iLambda\Im\msc E\)\Bigg]\rvt}{htF20}

\subsection{Remark on rigidity theorems}

%\lbl{odd:htrt}

We wish to demonstrate that rigidity theorems for the dragging potential and the generalised electrostatic potential as well as relations \eqref{nhtiqu0}, that hold for the Kerr-Newman solution, are preserved by the Harrison transformation to a MKN black hole. From equation \eqref{htomth} it follows 
\rov{\res{\pder{\omega}{\vartheta}}{N=0}=0\quad\imply\quad\res{\pder{\omega'}{\vartheta}}{N'=0}=0\rvs}{rtpres1}
so a MKN black hole satisfies the rigidity theorem for the dragging potential because the Kerr-Newman solution does. (Note that $N'=\left|\iLambda\right|N$ where $\left|\iLambda\right|>0$.)

Regarding the generalised electrostatic potential $\phi'$, let us first calculate its derivative from the definition \eqref{elstgen},
\rov{\pder{\phi'}{\vartheta}=-\pder{A_t'}{\vartheta}-\omega'\pder{A_\varphi'}{\vartheta}-\pder{\omega'}{\vartheta}A_\varphi'\rvt}{elstgenhtder}
Expressing coordinate components of the field strength tensor 
in terms of the tetrad component $F_{(\vartheta)(t)}$ and using \eqref{emernstgen}, we arrive at the following equation:
\rov{\pder{\phi}{\vartheta}=\frac{N\sqrt{g_{\vartheta\vartheta }}}{\sqrt{g_{\varphi\varphi}g_{rr}}}\pder{\iPhi'}{r}-\pder{\omega'}{\vartheta}A_\varphi'\rvc}{}
where $\iPhi'$ is given by \eqref{htphi}. 

Therefore
\rov{\res{\pder{\omega'}{\vartheta}}{N'=0}=0\quad\imply\quad\res{\pder{\phi'}{\vartheta}}{N'=0}=0\rvs}{}
MKN black holes satisfy the rigidity theorem for the generalised electrostatic potential as a consequence of the rigidity for the dragging potential.

Let us take the derivative of the equation \eqref{htomth} with respect to $r$
\drov{\smder{\omega'}{r}{\vartheta}=\pder{}{r}\(\left|\iLambda\right|^2\)\pder{\omega}{\vartheta}+\left|\iLambda\right|^2\smder{\omega}{r}{\vartheta}-\pder{}{r}\(N^2\)\frac{2}{N\sqrt{g_{rr}}}\sqrt{\frac{g_{\vartheta\vartheta}}{g_{\varphi\varphi}}}\(\Re\iLambda\pder{\Im\iLambda}{r}-\Im\iLambda\pder{\Re\iLambda}{r}\)-\zrov-N^2\pder{}{r}\[\frac{2}{N\sqrt{g_{rr}}}\sqrt{\frac{g_{\vartheta\vartheta}}{g_{\varphi\varphi}}}\(\Re\iLambda\pder{\Im\iLambda}{r}-\Im\iLambda\pder{\Re\iLambda}{r}\)\]\rvt}{}
Note that the product $N\sqrt{g_{rr}}$ must be finite and non-zero for $\sqrt{-g}$ non-degenerate. When we evaluate the resulting equation at $N=0$ (regarding \eqref{rtpres1}), we get
\rov{\res{\smder{\omega'}{r}{\vartheta}}{N'=0}=\res{\(\left|\iLambda\right|^2\smder{\omega}{r}{\vartheta}\)}{N=0}-\pder{}{r}\(N^2\)\res{\[\frac{2}{N\sqrt{g_{rr}}}\sqrt{\frac{g_{\vartheta\vartheta}}{g_{\varphi\varphi}}}\(\Re\iLambda\pder{\Im\iLambda}{r}-\Im\iLambda\pder{\Re\iLambda}{r}\)\]}{N=0}\rvt}{}
The radial derivative of $N^2$ vanishes on the degenerate horizon, so we can state
\rov{\res{\smder{\omega}{r}{\vartheta}}{N=0, \varkappa=0}=0\quad\imply\quad\res{\smder{\omega'}{r}{\vartheta}}{N'=0, \varkappa=0}=0\rvt}{}
This is the result we wanted to prove: The radial derivative of the dragging potential for the MKN black hole does not depend on $\vartheta$ when evaluated on the degenerate horizon because the same holds for the Kerr-Newman solution.

Similarly, we take the derivative of equation \eqref{elstgenhtder} with respect to $r$ to get
\rov{\smder{\phi'}{r}{\vartheta}=\pder{}{r}\(N^2\)\frac{1}{N\sqrt{g_{rr}}}\sqrt{\frac{g_{\vartheta\vartheta}}{g_{\varphi\varphi}}}\pder{\Im\iPhi'}{r}+N^2\pder{}{r}\(\frac{1}{N\sqrt{g_{rr}}}\sqrt{\frac{g_{\vartheta\vartheta}}{g_{\varphi\varphi}}}\pder{\Im\iPhi'}{r}\)-\smder{\omega'}{r}{\vartheta}A_\varphi'-\pder{\omega'}{\vartheta}\pder{A_\varphi'}{r}\rvt}{}
We restrict this equation to $N=0$, so that it reduces to the form
\rov{\res{\smder{\phi'}{r}{\vartheta}}{N'=0}=\pder{}{r}\(N^2\)\res{\(\frac{1}{N\sqrt{g_{rr}}}\sqrt{\frac{g_{\vartheta\vartheta}}{g_{\varphi\varphi}}}\pder{\Im\iPhi'}{r}\)}{N=0}-\res{\(\smder{\omega'}{r}{\vartheta}A_\varphi'\)}{N'=0}\rvc}{}
from which we see that
\rov{\res{\smder{\omega'}{r}{\vartheta}}{N'=0, \varkappa=0}=0\quad\imply\quad\res{\smder{\phi'}{r}{\vartheta}}{N'=0, \varkappa=0}=0\rvt}{}
Hence the radial derivative of the generalised electrostatic potential does not depend on $\vartheta$ on the degenerate horizon of an extremal MKN black hole.

\end{document}